\PassOptionsToPackage{numbers,sort&compress}{natbib}
\documentclass[preprint,12pt]{elsarticle}

\usepackage{amssymb}
\usepackage{amsmath}
\usepackage{subcaption}
\usepackage{float}
\usepackage{booktabs} 
\usepackage{hyperref}  
\usepackage{multirow}      %
\usepackage{algorithmic}
\usepackage{algorithm}

\usepackage{xcolor}
\newcommand{\added}[1]{\textcolor{black}{#1}}      
\newcommand{\modified}[1]{\textcolor{black}{#1}}    

\journal{Signal Processing}

\begin{document}
	
	\begin{frontmatter}
		
		

		\title{Target Localization with Coprime Multistatic MIMO Radar via Coupled Canonical Polyadic Decomposition Based on Joint Eigenvalue Decomposition}
		
		
		\author{Guo-Zhao Liao \fnref{label1}}
		\author{Xiao-Feng Gong \fnref{label1}}
		\author{Wei Liu \fnref{label2}}
		\author{Hing Cheung So \fnref{label3}}
		
		\affiliation[label1]{organization={School of Information and Communication Engineering},
			addressline={Dalian University of Technology},
			city={Dalian},
			postcode={116024},
			country={China}}
		
		\affiliation[label2]{organization={Department of Electrical and Electronic Engineering},
			addressline={The Hong Kong Polytechnic University},
			city={Hong Kong},
			country={China}}
		
		\affiliation[label3]{organization={Department of Electrical Engineering},
			addressline={City University of Hong Kong},
			city={Hong Kong},
			country={China}}
		
		\begin{abstract}		
			This paper investigates target localization using a multistatic multiple-input multiple-output (MIMO) radar system with two distinct coprime array configurations: coprime L-shaped arrays and coprime planar arrays. The observed signals are modeled as tensors that admit a coupled canonical polyadic decomposition (C-CPD) model. For each configuration, a C-CPD method is presented based on joint eigenvalue decomposition (J-EVD). This computational framework includes (semi-)algebraic and optimization-based C-CPD algorithms and target localization that fuses direction-of-arrivals (DOAs) information to calculate the optimal position of each target. Specifically, the proposed (semi-)algebraic methods exploit the rotational invariance of the Vandermonde structure in coprime arrays, similar to the multiple invariance property of \added{estimation of signal parameters via rotational invariance techniques} (ESPRIT), which transforms the model into a J-EVD problem and reduces computational complexity. The study also investigates the working conditions of the algorithm to understand model identifiability. Additionally, the proposed method does not rely on prior knowledge of non-orthogonal probing waveforms and is effective in challenging underdetermined scenarios. Experimental results demonstrate that our method outperforms existing tensor-based approaches in both accuracy and computational efficiency.
		\end{abstract}	
		
        \begin{graphicalabstract}
        	\begin{center}
	     	\includegraphics[width=5.5in]{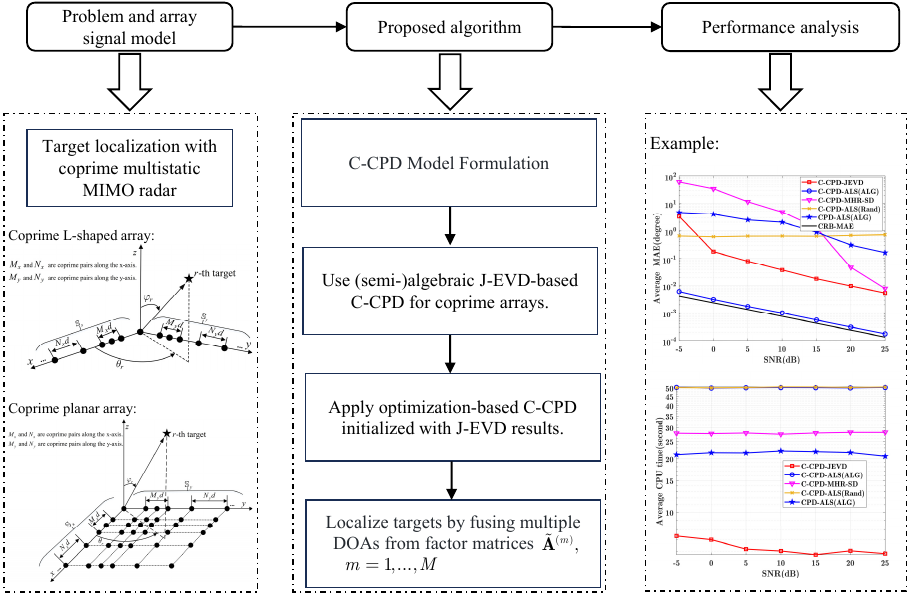}%
        	\end{center}
        \end{graphicalabstract}
		
		\begin{highlights}
		\item A target localization method is proposed for a multistatic MIMO radar system using two distinct coprime array configurations: coprime L-shaped arrays and coprime planar arrays, and model the observed data from multiple receive arrays using the C-CPD framework.
	
		\item A novel (semi-)algebraic C-CPD algorithm is developed, which leverages the Vandermonde structure in coprime arrays, transforming the C-CPD problem into a joint eigenvalue decomposition (J-EVD) problem and reducing computational complexity.
		
		\item Working conditions for the proposed J-EVD-based (semi-)algebraic C-CPD algorithm are derived for the two coprime array configurations, which are crucial for ensuring its effectiveness in MIMO radar applications.
		
		\item Extensive experiments are conducted, demonstrating the superiority of the proposed method in both accuracy and computational efficiency compared to existing tensor-based approaches.
		\end{highlights}
		
		
	\end{frontmatter}
	\section{Introduction}
	\label{sec1}

	\modified{Tensor-based techniques for multiple-input multiple-output (MIMO) radar processing have gained considerable attention in the past decade, primarily due to their ability to harness multilinear (ML) structures within the data.} This exploitation enhances target identification accuracy and resolution while eliminating the need for additional parameter pairing \cite{Nion2010, Chen2021, Sidiropoulos2017}. Concurrently, multistatic (MS) MIMO radar has also attracted increased research interest owing to its capacity to provide rich spatial diversity and improve target localization accuracy \cite{Nion2010, Liu2014, Chernyak2014, Gong2022, Liu2024, Fang2024, Liao2024}.
	
	\modified{Recent studies have also investigated tensor-based modeling techniques for MS MIMO radar, with a focus on exploiting array coupling and multilinear structures in the received data \cite{Nion2010, Gong2022, Liao2024}}. In signal processing, the key advancement from monostatic/\allowbreak bistatic to MS MIMO radar lies in the shift from single-array approaches to the fusion of data from multiple receive arrays, along with the leveraging of array coupling and the \modified{ML} structure of datasets using structured tensor techniques. For example, in \cite{Nion2010}, the matched signal for a multi-transmit, multi-receive, multi-pulse (MT-MR-MP) MIMO radar is modeled using a structured \modified{ML} rank-(\textit{Nr}, \textit{Mr}, $\cdot$) block term decomposition model to address subarray coupling. Similarly, \cite{Gong2022} applies coupled canonical polyadic decomposition (C-CPD) to a single-transmit, multi-receive (ST-MR) MIMO radar without requiring prior knowledge of probing waveforms. In \cite{Liao2024}, the matched signal for an MT-MR-MP MIMO radar is described using a double C-CPD (DC-CPD) model to exploit subarray double coupling. However, these studies primarily utilize the ML and coupling structures of MIMO radar signals without focusing on specific array geometries.
	
	\modified{More recently, array configurations like sparse arrays—including nested, coprime, and minimum redundancy arrays—have gained popularity for their ability to mitigate mutual coupling, offering an advantage over uniform linear arrays (see \cite{Vaidyanathan2011,Qin2015,Liu2016,Shen2016,Shi2017,Shi2017b,Zhou2018,Zheng2021,Zhang2023,Xu2024} and references therein).} The relationship between irregular sensor array configurations, such as sparse arrays, and C-CPD methods has been well-established \cite{Sorensen2016,Sorensen2017a,Sorensen2017b,Sorensen2018}. These studies advocate a shift from traditional CPD-based methods \cite{Sidiropoulos2000} to coupled tensor-based array processing, linking multidimensional harmonic retrieval (MHR) \cite{Haardt1998b} with multiple invariance estimation of signal parameters via rotational invariance technique (MI-ESPRIT) \cite{Swindlehurst1992}, enabling simultaneous consideration of harmonic structures. Specifically, in \cite{Sorensen2017a} and \cite{Sorensen2017b}, the proposed C-CPD framework for MHR supports multirate sampling algebraically.

	However, the limitations of previous algorithms are twofold: (1) existing coprime array related works are mainly based on “virtual array” in the higher-order statistics domain for aperture extension, which requires the sources to be uncorrelated (in our case of MIMO radar, they require uncorrelation for the radar cross section (RCS) coefficient vectors of different targets). Moreover, to accurately compute the statistics of the observed signals, large number of pulses within a coherent processing interval (CPI) are required to mitigate the so-called finite sampling effects. Hence, in case where the RCS coefficient vectors are correlated, or the number of pulses is small, the “virtual array” based works may be less effective. (2) existing C-CPD based works on array or MIMO radar processing mainly use unconstrained C-CPD methods. These methods work in the data domain and thus do not require the sources (or RCS coefficient vectors in the context of MIMO radar) to be uncorrelated. However, these techniques mostly ignore the potential harmonic structures in the steering vectors of a coprime array. 
		
	In this work, a target localization method is proposed for MS MIMO radar, employing two distinct coprime array configurations: coprime L-shaped arrays (CPLsAs) and coprime planar arrays (CPPAs). \added{These two configurations are considered to demonstrate the generality of the proposed method, as both contain local sparse uniform linear subarrays. These subarrays yield Vandermonde-structured steering matrices and induce spatial sampling patterns analogous to multirate sampling, which help improve target localization accuracy.} For each configuration, a corresponding C-CPD model is developed for an MS MIMO radar system with a single transmitter and multiple receivers. Specifically, we exploit the rotational invariance of the Vandermonde structure in the local sparse uniform linear subarrays (ULAs) of each configuration, transforming the C-CPD problem into a joint eigenvalue decomposition (J-EVD) problem, which reduces computational complexity. We also investigate the conditions necessary for the J-EVD-based C-CPD algorithm to ensure model identifiability and provide insights into its applicability. Finally, a post-processing approach is employed to calculate and fuse the direction-of-arrival (DOA) information for target localization.
	
	The main contributions of this paper is summarized as follows.
	\vspace{-6pt} 
	\begin{itemize}
	\item A target localization method is proposed for MS MIMO radar with two distinct coprime array configurations: CPLsAs and CPPAs, and a C-CPD model is established based on the observed data from multiple receive arrays.
	\vspace{-6pt} 
	\item A novel semi-algebraic C-CPD algorithm is developed based on J-EVD, which fully exploits the Vandermonde structure in coprime arrays to transform the C-CPD problem into a J-EVD problem, reducing computational complexity; the working conditions are also established to provide insights into model identifiability.
	\vspace{-6pt} 
	\item Extensive simulations are conducted to demonstrate the advantages of the proposed method over existing tensor-based approaches in scenarios involving both overdetermined and underdetermined cases.
	\end{itemize}
	\vspace{-5pt} 

The remainder of this paper is organized as follows. \modified{In Section~\ref{sec2}, we introduce the C-CPD modeling of two coprime MS array configurations: CPLsAs and CPPAs. Section~\ref{sec3} details the proposed method, including both (semi-)algebraic and optimization-based C-CPD algorithms, as well as DOA-based target localization for CPLsAs and CPPAs. The effectiveness of the proposed approach is demonstrated through numerical simulations in Section~\ref{sec4}. Finally, Section~\ref{sec5} concludes the paper.}

    \textit{Notations}: Scalars, vectors, matrices, and tensors are denoted by italic lowercase, \modified{bold lowercase, bold uppercase, and calligraphic uppercase letters}, respectively. The \(r\)-th column and the \((i,j)\)-th entry of matrix \(\mathbf{A}\) are denoted by \(\mathbf{a}_{r}\) and \(a_{ij}\), respectively. \(|\mathbb{S}|\) denotes the cardinality of a set \(\mathbb{S}\), \added{ \(\mathbb{S}_1 \cup \mathbb{S}_2\) denotes the union of two sets \(\mathbb{S}_1\) and \(\mathbb{S}_2\), and \(\mathbb{S}_1 \times \mathbb{S}_2\) denotes their Cartesian product}. We use \textsc{Matlab} notation to denote subtensors or submatrices obtained by fixing certain indices or index ranges of a tensor. For instance, \(\mathcal{T}(:,:,s)\) denotes the \(s\)-th frontal slice of tensor \(\mathcal{T}\) obtained by fixing the third index of \(\mathcal{T}\) to \(s\), and \(\mathbf{A}(1{:}K,:)\) represents the submatrix of \(\mathbf{A}\) consisting of the rows from 1 to \(K\). Similarly, \(\mathbf{A}(\mathbb{Q},:)\) denotes the submatrix of \(\mathbf{A}\) consisting of rows indexed by the elements of the set \(\mathbb{Q}\). The matrices \(\underline{\mathbf{A}}^{(1)}\) and \(\overline{\mathbf{A}}^{(1)} \in \mathbb{C}^{(I_1-1) \times R}\) are obtained by removing the last and first rows of \(\mathbf{A}^{(1)}\), respectively. The operators \modified{\((\cdot)^{\mathrm{T}}\)}, \((\cdot)^{-1}\), and \((\cdot)^{\dagger}\) denote the transpose, inverse, and Moore--Penrose pseudo-inverse, respectively. \added{The notations \(|\cdot|\), \(\|\cdot\|\), and \(\|\cdot\|_F\) represent the absolute value, Euclidean norm, and Frobenius norm, respectively}, while \(\circ\) and \(\odot\) denote the vector outer product and the Khatri--Rao product, respectively. \added{$\cap$ denotes the intersection operation between two subspaces.} We denote the mode-2 unfolding of a third-order tensor \(\mathcal{T} \in \mathbb{C}^{I \times J \times K}\) by \(\mathbf{T}_{(2)}\), defined such that \((\mathbf{T}_{(2)})_{(i-1)K + k, j} = \mathcal{T}_{i,j,k}\), and for a fourth-order tensor \(\mathcal{T} \in \mathbb{C}^{I \times J \times K \times L}\), its mode-3 unfolding is denoted by \(\mathbf{T}_{(3)}\), defined such that \((\mathbf{T}_{(3)})_{(i-1)JL + (j-1)L + l, k} = \mathcal{T}_{i,j,k,l}\).

	A CPD expresses \(\mathcal{T}\) as the sum of minimal number, \textit{R}, of rank-1 terms, and \textit{R} is defined as the rank of \(\mathcal{T}\):
	 \vspace{-5pt} 
	    \begin{equation}
			{{\mathcal{T}}}={{[\![{{\mathbf{A}}},\mathbf{B},{{\mathbf{C}}}]\!]}_{R}}\triangleq \sum\limits_{r=1}^{R}{\mathbf{a}_{r}\circ {{\mathbf{b}}_{r}}\circ \mathbf{c}_{r}}.
		\end{equation}
		
	A C-CPD writes a set of tensors \(\text{ }\!\!\{\!\!\text{ }{{\mathcal{T}}^{(m)}},m=1,...,M\}\) as the sum of minimal number of coupled rank-1 terms:
	\vspace{-5pt} 
		\begin{equation}
			{{\mathcal{T}}^{(m)}}={{[\![{{\mathbf{A}}^{(m)}},\mathbf{B},{{\mathbf{C}}^{(m)}}]\!]}_{R}}\triangleq \sum\limits_{r=1}^{R}{\mathbf{a}_{r}^{(m)}\circ {{\mathbf{b}}_{r}}\circ \mathbf{c}_{r}^{(m)}},
		\end{equation}where \(\textit{R}\) is defined as the coupled rank of \(\{{{\mathcal{T}}^{(m)}}\}\).
		
	\section{Coprime MS Array Signal Model}
	\label{sec2}		
	This section describes the development of the coprime MS array signal model, focusing on the sensor locations for CPLsAs and CPPAs, and the data model and assumptions for the MS system.
	
	 \subsection{Sensor locations for CPLsAs}	
		Consider an MS MIMO radar system with a single transmit array and multiple receive arrays. All receive arrays are CPLsA, while the transmit array is a CPLsA but could be any configuration. For convenience, we label the MS MIMO radar with CPLsA as MS CPLsA. Specifically, the $m$-th receive array features two sparse ULAs in both $x$- and $y$-axes. In the $x$-axis, the first subarray has $I_{x,1}^{(m)} $ sensors with inter-element spacing $M_{x}^{(m)}d$ and the second subarray has $I_{x,2}^{(m)}$ sensors with inter-element spacing $N_{x}^{(m)}d$, where $M_{x}^{(m)}$ and $N_{x}^{(m)}$ are coprime integers, and $d=\lambda /2$, with $\lambda $ being the signal wavelength. Due to their coprime property, these subarrays overlap at a reference element located at $(0,0,0)$. The sensor locations are given by $\mathbb{S}_{x}^{(m)}d$, where $\mathbb{S}_{x}^{(m)} $ is an integer set defined as $\mathbb{S}_{x}^{(m)}=\{({{l}_{x,i}},0,0,)|i=1,2,\ldots ,|\mathbb{S}_{x}^{(m)}|\} $, which can be decomposed as $\mathbb{S}_{x}^{(m)}=\mathbb{S}_{x,1}^{(m)}\cup \mathbb{S}_{x,2}^{(m)} $, where $\mathbb{S}_{x,1}^{(m)}=\{(M_{x}^{(m)}{{m}_{x}},0,0,)|0\le {{m}_{x}}\le I_{x,1}^{(m)}-1\}$, and $\mathbb{S}_{x,2}^{(m)}=\{(N_{x}^{(m)}{{n}_{x}},0,0)|0\le {{n}_{x}}\le I_{x,2}^{(m)}-1\}$. 
		
		Similarly, the $y$-axis of the $m$-th receive array consists of $|\mathbb{S}_{y}^{(m)}|$ sensors. The sensor locations are given by $\mathbb{S}_{y}^{(m)}d$, where $\mathbb{S}_{y}^{(m)} $ is an integer set defined as $\mathbb{S}_{y}^{(m)}=\{(0,{{l}_{y,i}},0)|i=1,2,\ldots ,|\mathbb{S}_{y}^{(m)}|\} $, which can be expressed as $\mathbb{S}_{y}^{(m)}=\mathbb{S}_{y,1}^{(m)}\cup \mathbb{S}_{y,2}^{(m)} $, where $\mathbb{S}_{y,1}^{(m)}=\{(0,M_{y}^{(m)}{{m}_{y}},0)|{{m}_{y}}=0,1,...,I_{y,1}^{(m)}-1\},$ and $\mathbb{S}_{y,2}^{(m)}=\{(0,N_{y}^{(m)}{{n}_{y}},0)|{{n}_{y}}=0,1,...,I_{y,2}^{(m)}-1\}$.
			
		Consequently, the total number of elements in the $m$-th receive array is $|{{\mathbb{S}}^{(m)}}|=|\mathbb{S}_{x}^{(m)}\cup \mathbb{S}_{y}^{(m)}|={{I}^{(m)}}$, and the sensor locations are given by $\mathbb{S}_{x}^{(m)}d$, where ${{\mathbb{S}}^{(m)}}$ is an integer set defined as ${{\mathbb{S}}^{(m)}}=\mathbb{S}_{x}^{(m)}\cup \mathbb{S}_{y}^{(m)}$.	
		
\begin{figure}[t]
		\centering
		\includegraphics[width=0.95\textwidth]{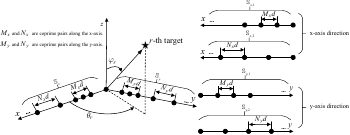}
		\caption{The geometry of the designed CPLsA (Left) and its local sparse ULAs (Right).}\label{fig1CPLsA}
		\end{figure}

		For clarity, the geometry of the designed CPLsA and its corresponding local sparse ULAs are illustrated in Fig. \ref{fig1CPLsA}.	
			
		\subsection{Sensor locations for CPPAs}	
		Consider an MS MIMO radar system with a single transmit array and multiple receive arrays. All receive arrays are CPPAs, while the transmit array is a CPPA but could be any configuration. For convenience, we label the MS MIMO radar with CPPA as MS CPPA. The CPPA in the \(m\)-th receive array is constructed from the Cartesian product of the sensor location sets along the \(x\)-axis and \(y\)-axis. Specifically, the sensor location set along the \(x\)-axis, denoted by \(\mathbb{S}_x^{(m)}\), is composed of two sparse ULAs, $\mathbb{S}_{x,1}^{(m)}$ and $\mathbb{S}_{x,2}^{(m)}$. The first subarray contains $I_{x,1}^{(m)}$ sensors with an inter-element spacing of $M_x^{(m)}d$, where $d = \lambda/2$ and $\lambda$ is the signal wavelength, and is given by the sensor location set $\mathbb{S}_{x,1}^{(m)} = \{(M_x^{(m)}m_x, 0, 0) \mid m_x = 0, 1, \dots, I_{x,1}^{(m)} - 1\}$. The second subarray contains $I_{x,2}^{(m)}$ sensors with an inter-element spacing of $N_x^{(m)}d$, given by the sensor location set $\mathbb{S}_{x,2}^{(m)} = \{(N_x^{(m)}n_x, 0, 0) \mid n_x = 0, 1, \dots, I_{x,2}^{(m)} - 1\}$, where $M_x^{(m)}$ and $N_x^{(m)}$ are coprime integers. Thus, the overall sensor location set along the \(x\)-axis is formed by combining the two subarrays, $\mathbb{S}_x^{(m)} = \mathbb{S}_{x,1}^{(m)} \cup \mathbb{S}_{x,2}^{(m)}$. Due to the coprime property of $M_x^{(m)}$ and $N_x^{(m)}$, these two subarrays overlap at a reference element located at $(0,0,0)$.
		
		Similarly, the sensor location set along the \(y\)-axis, denoted by $\mathbb{S}_y^{(m)}$, is also formed by combining two sparse ULAs, $\mathbb{S}_{y,1}^{(m)}$ and $\mathbb{S}_{y,2}^{(m)}$. The first subarray $\mathbb{S}_{y,1}^{(m)}$ contains $I_{y,1}^{(m)}$ sensors with an inter-element spacing of $M_y^{(m)}d$, given by the sensor location set $\mathbb{S}_{y,1}^{(m)} = \{(0, M_y^{(m)}m_y, 0) \mid m_y = 0, 1, \dots, I_{y,1}^{(m)} - 1\}$. The second subarray $\mathbb{S}_{y,2}^{(m)}$ contains $I_{y,2}^{(m)}$ sensors with an inter-element spacing of $N_y^{(m)}d$, given by the sensor location set $\mathbb{S}_{y,2}^{(m)} = \{(0, N_y^{(m)}n_y, 0) \mid n_y = 0, 1, \dots, I_{y,2}^{(m)} - 1\}$, where $M_y^{(m)}$ and $N_y^{(m)}$ are coprime integers. Thus, the overall sensor location set along the \(y\)-axis is formed by combining the two subarrays, $\mathbb{S}_y^{(m)} = \mathbb{S}_{y,1}^{(m)} \cup \mathbb{S}_{y,2}^{(m)}$, and these two subarrays also overlap at a reference element located at the $(0,0,0)$.
		
		For simplicity, we set $M_y^{(m)} = M_x^{(m)} = M'$ and $N_y^{(m)} = N_x^{(m)} = N'$ for all $m = 1, \dots, M$. Consequently, the complete CPPA in the $m$-th receive array is given by the Cartesian product of the sensor location sets along the $x$-axes and $y$-axes, $\mathbb{S}^{(m)} = \mathbb{S}_x^{(m)} \times \mathbb{S}_y^{(m)} = \{(l_{x,i}, l_{y,j}, 0) \mid l_{x,i} \in \mathbb{S}_x^{(m)}, l_{y,j} \in \mathbb{S}_y^{(m)}\}$.
		
\begin{figure}[t]
	    	\centering
	    	\includegraphics[width=0.95\textwidth]{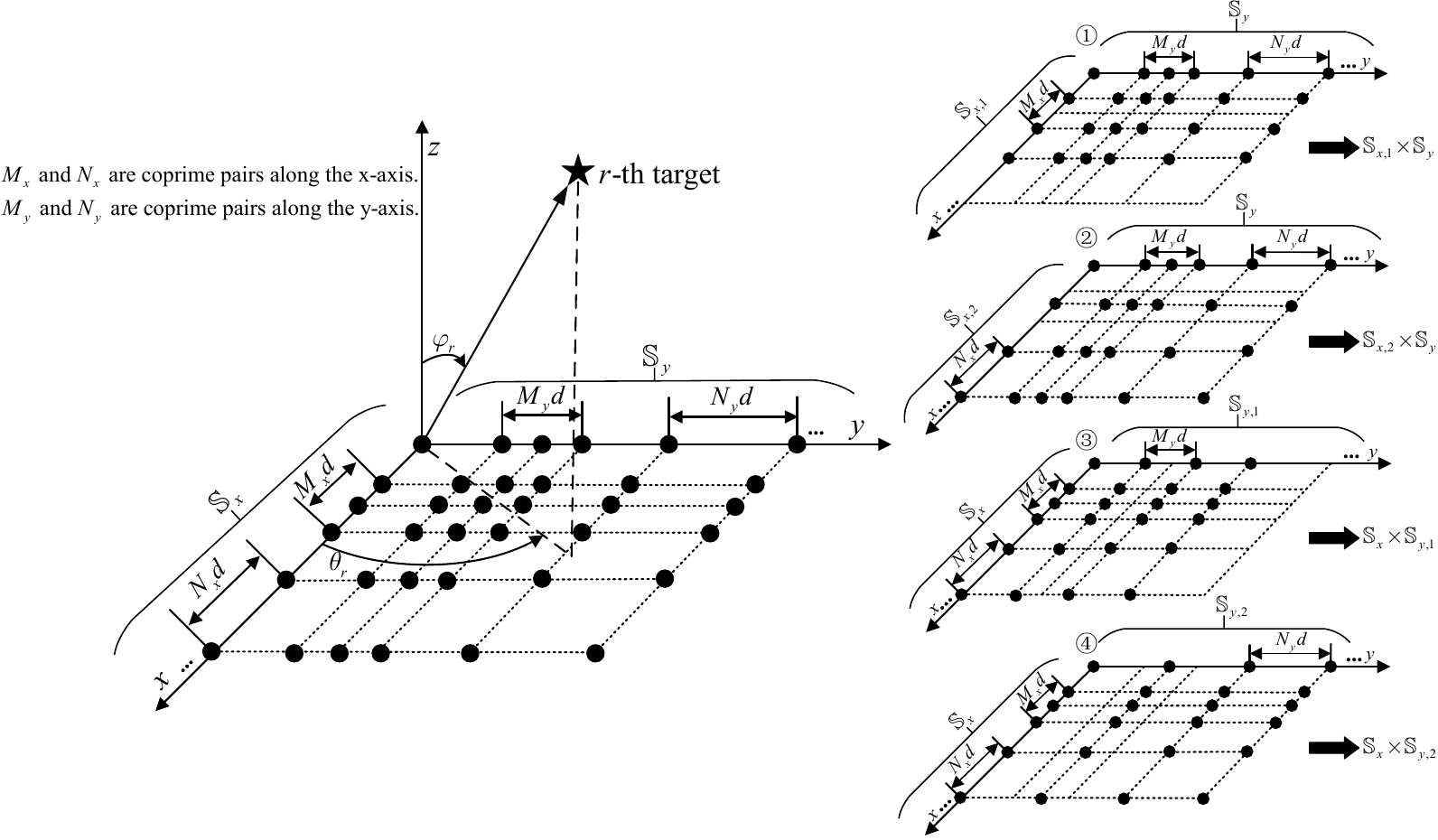}
	    	\caption{The geometry of the designed CPPA (Left) and its four types of local sparse UPAs (Right).} \label{fig2CPPAs}
	    \end{figure}

		Specifically, both the \(x\)- and \(y\)-axes consist of two sparse ULAs, and their Cartesian product forms four types of subarrays. For clarity, the geometry of the CPPA and its four types of local sparse uniform planar subarrays (UPAs) is shown in Fig. \ref{fig2CPPAs}.
	
		\subsection{Data model and assumptions}
	
	    We assume: (A1) the targets are located in the far-field with regard to both transmit and receive arrays; (A2) transmitted and received signals are narrowband; (A3) multiple pulses are emitted in each CPI; (A4) the RCS coefficients of distinct targets vary independently from pulse to pulse (Swerling II model \cite{Swerling1960}); (A5) direction of departures (DODs) and DOAs for different targets are different; (A6) there is no angle ambiguity in each transmit and receive array.
	    
	    The output signal of the $m$-th receive array during the $k$-th pulse period can be expressed as (the noise term is omitted for convenience):
	    \vspace{-5pt} 
	    \begin{equation}
	    	\mathbf{X}_{k}^{(m)} = \sum\nolimits_{r=1}^{R} c_{k,r}^{(m)} \mathbf{a}_{r}^{(m)} \mathbf{t}_{r}^{\modified{\mathrm{T}}} {\mathbf{S}}^{\modified{\mathrm{T}}} \in \mathbb{C}^{{I^{(m)}} \times T},
	    \label{eq:Xkm}
	    \end{equation}
	  where \(\mathbf{S} \in \mathbb{C}^{T \times J}\) contains the probing signals from the transmit array, with each column representing a signal sampled during one pulse period, and \(T\) denotes the number of samples per pulse. \added{The term \(c_{k,r}^{(m)}\) denotes the RCS coefficient of the \(r\)-th target with respect to the \(m\)-th receive array during the \(k\)-th pulse period, with \(r = 1,\dots,R\), \(k = 1,\dots,K\), and \(m = 1,\dots,M\); here, \(R\), \(K\), and \(M\) denote the numbers of targets, pulse periods, and receive arrays, respectively.} The receive and transmit steering vectors \(\mathbf{a}_{r}^{(m)} \in \mathbb{C}^{I^{(m)}}\) and \(\mathbf{t}_r \in \mathbb{C}^{J}\) are defined as:
	    
        \vspace{-5pt} 
	   	\begin{equation}
		\left\{
		\begin{aligned}
			 \mathbf{a}_{r}^{(m)} &\triangleq \exp (\operatorname{i} 2\pi \lambda^{-1} [ \mathbf{i}_1^{(m)} \mathbf{v}_{(\theta_r^{(m)}, \varphi_r^{(m)})}, \ldots, \mathbf{i}_{{I}^{(m)}}^{(m)} \mathbf{v}_{( \theta_r^{(m)}, \varphi_r^{(m)})}]^{\modified{\mathrm{T}}}), \\
			 \mathbf{t}_r &\triangleq \exp (\operatorname{i} 2\pi \lambda^{-1} [ \mathbf{j}_1 \mathbf{v}_{\left( \alpha_r, \beta_r \right)}, \ldots, \mathbf{j}_J \mathbf{v}_{\left(\alpha_r, \beta_r \right)} ]^{\modified{\mathrm{T}}}),
		\end{aligned}  
		\right.
		\end{equation}where $\operatorname{i}$ is the imaginary unit, $J$ is the number of antennas in the transmit array, and ${{I}^{(m)}}$ the number in the $m$-th receive array; ${{\mathbf{v}}_{(\theta _{r}^{(m)},\varphi _{r}^{(m)})}} $ and ${{\mathbf{v}}_{({{\alpha }_{r}},{{\beta }_{r}})}}$ denote the DOA of the signal from the $r$-th target to the $m$-th receive array and the DOD of signal from the transmit array to the $r$-th target, respectively. Row vectors ${{\mathbf{j}}_{j}},\mathbf{i}_{i}^{(m)}\in {{\mathbb{R}}^{3}}$ represent the sensor locations in the transmit and receive arrays, respectively. Specifically, $\mathbf{i}_{i}^{(m)}\in {{\mathbb{S}}^{(m)}},i=1,2,...,|{{\mathbb{S}}^{(m)}}|$ indicates the position of the $i$-th sensor in the coprime array. The transmit array is configured as a coprime array, with a CPLsA for MS CPLsA and a CPPA for MS CPPA. 
	
     	Denote ${{\mathbf{b}}_{r}}\triangleq \mathbf{S}{{\mathbf{t}}_{r}}\in {{\mathbb{C}}^{T}}$, and then by \eqref{eq:Xkm} we obtain $x_{i,t,k}^{(m)}=\sum\nolimits_{r=1}^{R}{a_{i,r}^{(m)}{{b}_{t,r}}}c_{k,r}^{(m)}.$
	
    	Stacking matrices $\mathbf{X}_{k}^{(m)}$ for fixed $m$ and varying $k$ along the third mode to construct a third-order tensor ${{({{\mathcal{X}}^{(m)}})}_{:,:,k}}\triangleq \mathbf{X}_{k}^{(m)}$, we have:
        \vspace{-5pt} 
    	\begin{equation}
    		\mathcal{X}^{(m)} = [\![ \mathbf{A}^{(m)}, \mathbf{B}, \mathbf{C}^{(m)} ]\!]_{R},
    		\label{eq:Xm}
    	\end{equation}where ${{\mathbf{A}}^{(m)}}\triangleq [\mathbf{a}_{1}^{(m)},...,\mathbf{a}_{R}^{(m)}]$, $\mathbf{B}\triangleq [{{\mathbf{b}}_{1}},...,{{\mathbf{b}}_{R}}]$ and ${{\mathbf{C}}^{(m)}}\triangleq [\mathbf{c}_{1}^{(m)},...,\mathbf{c}_{R}^{(m)}]$. The set of tensors $\{{{\mathcal{X}}^{(m)}},m=1,...,M\}$ together admits a C-CPD.
              
A case is defined as underdetermined if the number of sensors is less than the number of targets, and overdetermined otherwise. Under the assumption that the factor matrix $\mathbf{B}$ has full column rank, the DOD estimation is always overdetermined, whereas the underdetermined condition primarily occurs in DOA estimation.

	    \section{Proposed Method}
	    \label{sec3}
	     Now we present a target localization framework for a coprime MS MIMO radar using C-CPD based on J-EVD. The overall process consists of two main stages: the computation of C-CPD based on J-EVD and target localization. 
	
	    \subsection{(Semi-) Algebraic C-CPD based on J-EVD for MS CPLsA}
	    \label{subsec3.1}
     An algorithm is termed algebraic if it relies solely on arithmetic operations, overdetermined linear equations, singular value decomposition (SVD), and generalized EVD (GEVD). If some operations are replaced by iterative processes for greater accuracy, it is referred to as (semi-)algebraic. For the C-CPD of a third-order tensor in \eqref{eq:Xm}, we assume that the second factor matrix \(\mathbf{B}\) has dimensions \(R \times R\) via dimensionality reduction \cite{Gong2022}, \cite{Gong2018}, \cite{Gong2019}. In the J-EVD based C-CPD algorithm for \added{MS} CPLsA, the Vandermonde structure in the first factor matrix, corresponding to the local sparse ULA configuration of CPLsA, is utilized to construct target matrices.
     
     	The proposed algorithm consists of the following steps:   	
     	
\noindent     	\textbf{Step 1: Construct} $\mathbf{G}_{v}^{(m)}\in {{\mathbb{C}}^{R\times R}}$ \textbf{for fixed} $m$ \textbf{and} $v=1,\ldots ,4$.  	

 To begin, we define the following notations. The factor matrix $\mathbf{A}^{(m)}$, corresponding to the steering vector of the $m$-th receive array in the MS CPLsA, is composed of the factor matrices $\mathbf{A}_{x}^{(m)}$ and $\mathbf{A}_{y}^{(m)}$, which represent the steering vectors corresponding to the $x$-axis and $y$-axis components, respectively. These matrices are submatrices of the steering matrix $\mathbf{A}^{(m)}$, with $\mathbf{A}_{x}^{(m)}$ of dimensions $|\mathbb{S}_{x}^{(m)}| \times R$ and $\mathbf{A}_{y}^{(m)}$ of dimensions $|\mathbb{S}_{y}^{(m)}| \times R$, where $\mathbf{A}_{x}^{(m)} = \mathbf{A}^{(m)}(\mathbb{Q}_{x}^{(m)},:)$ and $\mathbf{A}_{y}^{(m)} = \mathbf{A}^{(m)}(\mathbb{Q}_{y}^{(m)},:)$, with $\mathbb{Q}_{x}^{(m)}$ and $\mathbb{Q}_{y}^{(m)}$ being index sets within the set $\mathbb{S}^{(m)}$, corresponding to $\mathbb{S}_{x}^{(m)}$ and $\mathbb{S}_{y}^{(m)}$, respectively. For the two sparse ULAs along the $x$-axis, the Vandermonde matrices are defined as $\mathbf{A}_{x,1}^{(m)} = \mathbf{A}_{x}^{(m)}(\mathbb{Q}_{x,1}^{(m)},:)$ and $\mathbf{A}_{x,2}^{(m)} = \mathbf{A}_{x}^{(m)}(\mathbb{Q}_{x,2}^{(m)},:)$, where $\mathbb{Q}_{x,1}^{(m)}$ and $\mathbb{Q}_{x,2}^{(m)}$ are index sets within $\mathbb{S}_{x}^{(m)}$ corresponding to $\mathbb{S}_{x,1}^{(m)}$ and $\mathbb{S}_{x,2}^{(m)}$, respectively. Similarly, for the $y$-axis, the Vandermonde matrices are given by $\mathbf{A}_{y,1}^{(m)} = \mathbf{A}_{y}^{(m)}(\mathbb{Q}_{y,1}^{(m)},:)$ and $\mathbf{A}_{y,2}^{(m)} = \mathbf{A}_{y}^{(m)}(\mathbb{Q}_{y,2}^{(m)},:)$, with $\mathbb{Q}_{y,1}^{(m)}$ and $\mathbb{Q}_{y,2}^{(m)}$ representing the index sets within $\mathbb{S}_{y}^{(m)}$ corresponding to $\mathbb{S}_{y,1}^{(m)}$ and $\mathbb{S}_{y,2}^{(m)}$, respectively. Next, we construct $\mathbf{G}_{1}^{(m)}$ by leveraging the Vandermonde structure in $\mathbf{A}_{x,1}^{(m)}$. First, we need to extract the data tensor corresponding to the $x$-axis sensor elements of the $m$-th receive array, which can be expressed as follows:
    \vspace{-5pt} 
	\begin{equation}
		\mathcal{T}_{x}^{(m)} = \mathcal{T}^{(m)} ( \mathbb{Q}_{x}^{(m)},:,:) = [\![ \mathbf{A}_{x}^{(m)}, \mathbf{B}, \mathbf{C}^{(m)} ]\!]_{R}.
	\label{eq:Txm}
	\end{equation}
	
	Then, we extract the data tensor corresponding to the first sparse ULA along the $x$-axis:
    \vspace{-5pt} 
	\begin{equation}
		\mathcal{T}_{x,1}^{(m)}\triangleq\mathcal{T}_{x}^{(m)}(\mathbb{Q}_{x,1}^{(m)},:,:)={{[\![\mathbf{A}_{x,1}^{(m)},\mathbf{B},{{\mathbf{C}}^{(m)}}]\!]}_{R}}.
		\label{eq:Tx1m}
	\end{equation}
	
	After that, since $\mathcal{T}_{x,1}^{(m)}$ admits a third-order CPD \eqref{eq:Tx1m}, its mode-2 matrix representation $\mathbf{T}_{x,1}^{(m)}$ is:
	\vspace{-5pt} 
	\begin{equation}
	\mathbf{T}_{x,1}^{(m)}=(\mathbf{A}_{x,1}^{(m)}\odot {{\mathbf{C}}^{(m)}})\cdot {{\mathbf{B}}^{\modified{\mathrm{T}}}}.
		\label{eq:TTx1m}
	\end{equation}
	
	We choose two submatrices $\mathbf{M}_{x,1}^{(m,1)}$ and $\mathbf{M}_{x,1}^{(m,2)}$ from $\mathbf{T}_{x,1}^{(m)}$, each having dimensions $(I_{x,1}^{(m)}-1)K\times R$:
	\vspace{-5pt} 
\begin{equation}
	\left\{
	\begin{aligned}
		\mathbf{M}_{x,1}^{(m,1)} &\triangleq \mathbf{T}_{x,1}^{(m)}(1:(I_{x,1}^{(m)}-1)K,:) = (\underline{\mathbf{A}}_{x,1}^{(m)} \odot \mathbf{C}^{(m)}) \cdot \mathbf{B}^{\modified{\mathrm{T}}}, \\
		\mathbf{M}_{x,1}^{(m,2)} &\triangleq \mathbf{T}_{x,1}^{(m)}(K+1:I_{x,1}^{(m)}K,:) = (\overline{\mathbf{A}}_{x,1}^{(m)} \odot \mathbf{C}^{(m)}) \cdot \mathbf{B}^{\modified{\mathrm{T}}}.
	\end{aligned}
	\right.
	\label{eq:Mx1Mx2}
\end{equation}

As $\mathbf{A}_{x,1}^{(m)}$ is a Vandermonde matrix, we have $\overline{\mathbf{A}}_{x,1}^{(m)}=\underline{\mathbf{A}}_{x,1}^{(m)}\cdot \mathbf{Z}_{x,1}^{(m)}$, where $\mathbf{Z}_{x,1}^{(m)}$ is a diagonal matrix holding the Vandermonde generators of $\mathbf{A}_{x,1}^{(m)}$, $z_{x1,1}^{(m)},...,z_{x1,R}^{(m)}$, in its main diagonal. Therefore, we have the following result:
\vspace{-5pt} 
\begin{equation}
\overline{\mathbf{A}}_{x,1}^{(m)}\odot {\mathbf{C}^{(m)}}=(\underline{\mathbf{A}}_{x,1}^{(m)}\odot {\mathbf{C}^{(m)}})\cdot \mathbf{Z}_{x,1}^{(m)}.
	\label{eq:Ax1mC}
\end{equation}

Assume that $\mathbf{M}_{x,1}^{(m,1)}$ has full column rank, and construct the target matrix as $\mathbf{G}_{1}^{(m)}\triangleq {{[{{(\mathbf{M}_{x,1}^{(m,1)})}^{\dagger}}\mathbf{M}_{x,1}^{(m,2)}]^{\modified{\mathrm{T}}}}}\in {{\mathbb{C}}^{R\times R}}$. Then, after simple derivations, we obtain:
\vspace{-5pt} 
\begin{equation}
	\mathbf{G}_{1}^{(m)} \triangleq [(\mathbf{M}_{x,1}^{(m,1)})^\dagger \mathbf{M}_{x,1}^{(m,2)}]^{\modified{\mathrm{T}}} = \mathbf{B} \cdot \mathbf{Z}_{x,1}^{(m)} \cdot \mathbf{B}^{-1}.
	\label{eq:G1m}
\end{equation}

Next, we follow similar procedures to exploit the Vandermonde structure in $\mathbf{A}_{x,2}^{(m)}$, $\mathbf{A}_{y,1}^{(m)}$ and $\mathbf{A}_{y,2}^{(m)}$ to construct matrices $\mathbf{G}_{2}^{(m)},\mathbf{G}_{3}^{(m)}$ and $\mathbf{G}_{4}^{(m)}$, respectively. For instance, for the construction of $\mathbf{G}_{2}^{(m)}$, we replace $\mathbb{Q}_{x,1}^{(m)}$ in \eqref{eq:Tx1m} with $\mathbb{Q}_{x,2}^{(m)}$ to obtain the data tensor $\mathcal{T}_{x,2}^{(m)}$ corresponding to the $x$-axis direction of the $m$-th receive array. Then, $\mathbf{T}_{x,2}^{(m)}$ is obtained by \eqref{eq:TTx1m}. The submatrices of $\mathbf{T}_{x,2}^{(m)}$, namely $\mathbf{M}_{x,2}^{(m,1)}$ and $\mathbf{M}_{x,2}^{(m,2)}$, are constructed by \eqref{eq:Mx1Mx2}. Subsequently, $\mathbf{G}_{2}^{(m)}$ is computed with $\mathbf{M}_{x,2}^{(m,1)}$ and $\mathbf{M}_{x,2}^{(m,2)}$ by \eqref{eq:G1m}. Next, we replace \(\mathbb{Q}_{x}^{(m)}\) with \(\mathbb{Q}_{y}^{(m)}\) in \eqref{eq:Txm}, obtaining the tensor \(\mathcal{T}_{y}^{(m)}\). Subsequently, we utilize the index sets \(\mathbb{Q}_{y,1}^{(m)}\) and \(\mathbb{Q}_{y,2}^{(m)}\) to extract the corresponding data tensor from \(\mathcal{T}_{y}^{(m)}\) and follow a similar process as described in \eqref{eq:Tx1m}--\eqref{eq:G1m} to \(\mathbf{G}_{3}^{(m)}\) and \(\mathbf{G}_{4}^{(m)}\). As a result, the matrices \(\mathbf{G}_{1}^{(m)}, \mathbf{G}_{2}^{(m)}, \mathbf{G}_{3}^{(m)}\), and \(\mathbf{G}_{4}^{(m)}\) can be collectively expressed in the following J-EVD formulation:
\vspace{-5pt} 
\begin{equation}
	\left\{
	\begin{aligned}
		\mathbf{G}_{1}^{(m)} &= \mathbf{B} \cdot \mathbf{Z}_{x,1}^{(m)} \cdot \mathbf{B}^{-1}, \quad \mathbf{G}_{2}^{(m)} = \mathbf{B} \cdot \mathbf{Z}_{x,2}^{(m)} \cdot \mathbf{B}^{-1}, \\
		\mathbf{G}_{3}^{(m)} &= \mathbf{B} \cdot \mathbf{Z}_{y,1}^{(m)} \cdot \mathbf{B}^{-1}, \quad \mathbf{G}_{4}^{(m)} = \mathbf{B} \cdot \mathbf{Z}_{y,2}^{(m)} \cdot \mathbf{B}^{-1},
	\end{aligned}
	\right.
	\label{eq:G1G2G3G4}
\end{equation}
where \(\mathbf{Z}_{x,1}^{(m)}, \mathbf{Z}_{x,2}^{(m)}, \mathbf{Z}_{y,1}^{(m)}\), and \(\mathbf{Z}_{y,2}^{(m)}\) are diagonal matrices containing the Vandermonde generators of \(\mathbf{A}_{x,1}^{(m)}, \mathbf{A}_{x,2}^{(m)}, \mathbf{A}_{y,1}^{(m)}\), and \(\mathbf{A}_{y,2}^{(m)}\) on their main diagonals, respectively. Note that in this study, the goal of J-EVD is to find an invertible matrix \(\mathbf{B}\) such that \({{\mathbf{B}}^{-1}} \cdot \mathbf{G}_{v}^{(m)} \cdot \mathbf{B}\) is diagonal for \(v \in \{1,2,3,4\}\).

\textit{Remark 1}: It should be noted that at least one of the matrices $\mathbf{M}_{x,1}^{(m,1)}, \allowbreak \mathbf{M}_{x,2}^{(m,1)}, \allowbreak \mathbf{M}_{y,1}^{(m,1)}$, and $\mathbf{M}_{y,2}^{(m,1)}$, for $m = 1, \dots, M$, is required to have full column rank. This ensures that at least one target matrix is available for J-EVD (if only one target matrix is successfully constructed, J-EVD reduces to EVD). Generically\footnote{A property is called generic if it holds with probability one in the Lebesgue measure.}, this is equivalent to requiring the following working conditions for the proposed J-EVD based C-CPD algorithm:
\vspace{-5pt} 
\begin{equation}
	\left\{
	\begin{aligned}
		& \min (T, J) \geq R, \\
		& (I' - 1) K \geq R,
	\end{aligned}
	\right.
\end{equation}
where $I'$ represents the maximum value among the elements in the set $\{I_{x,1}^{(m)},\allowbreak I_{x,2}^{(m)},\allowbreak I_{y,1}^{(m)},\allowbreak I_{y,2}^{(m)}, m = 1, \dots, M\}$. These conditions can be regarded as generic sufficient uniqueness conditions for the C-CPD problem in \eqref{eq:Xm} of the proposed J-EVD based method for the MS CPLsA based MIMO radar structure.

\noindent     	\textbf{Step 2: Compute} $\mathbf{B}$ \textbf{via J-EVD of} $\{\mathbf{G}_{v}^{(m)},v=1,...,4,m=1,...,M\}$, \textbf{and} ${{\mathbf{A}}^{(m)}}$ \textbf{and} ${{\mathbf{C}}^{(m)}}$ \textbf{using rank-1 approximation}.

Note that \eqref{eq:G1G2G3G4} implies that, for all $m$, the matrices $\{\mathbf{G}_{v}^{(m)}, v=1, \allowbreak \dots,4, m=1,\dots,M\}$ collectively admit a J-EVD. The computation of the J-EVD yields the factor matrix $\mathbf{B}$ and the generators of $\{\mathbf{A}_{x,1}^{(m)}, \mathbf{A}_{x,2}^{(m)}, \mathbf{A}_{y,1}^{(m)},$ and $\mathbf{A}_{y,2}^{(m)} \allowbreak, m=1,\dots,M\}$. Notably, employing J-EVD to simultaneously extract these generators without automatic pairing, reduces computational complexity. This algorithm formulates the J-EVD problem as a structured CPD problem:
\vspace{-5pt} 
\begin{equation}
	\mathcal{G} = {[\![\mathbf{B}, \mathbf{D}, \mathbf{F}]\!]}_{R},\ \text{s.t.},\ \mathbf{B} \cdot \mathbf{D}^{\modified{\mathrm{T}}} = \mathbf{I},
	\label{eq:daG}
\end{equation}
where
\vspace{-5pt} 
\begin{equation}
	\left\{
	\begin{aligned}
	&	\mathcal{G}_{(:,:,w)} \triangleq \mathbf{G}'_{w}, \quad w \in \{1,2,3,\dots,4M\}, \\
	&	\mathbf{F} \triangleq [ \operatorname{diag}(\mathbf{Z}^{(1)}), \operatorname{diag}(\mathbf{Z}^{(2)}), \dots, \operatorname{diag}(\mathbf{Z}^{(M)}) ]^{\modified{\mathrm{T}}},
	\end{aligned}
	\right.
\end{equation}with ${{\mathbf{Z}}^{(m)}}=[\operatorname{diag}(\mathbf{Z}_{x,1}^{(m)}),\operatorname{diag}(\mathbf{Z}_{x,2}^{(m)}),\operatorname{diag}(\mathbf{Z}_{y,1}^{(m)}),\operatorname{diag}(\mathbf{Z}_{y,2}^{(m)})]$ and ${{\mathbf{{G}'}}_{4(m-1)+v}}=\mathbf{G}_{v}^{(m)},v=1,...,4,m=1,...,M.$

Then, the structured CPD \eqref{eq:daG} can be further formulated as a regularized LS problem: 
\vspace{-5pt} 
\begin{equation}
	\{\tilde{\mathbf{B}}, \tilde{\mathbf{D}}, \tilde{\mathbf{F}}\} = \underset{\mathbf{B}, \mathbf{D}, \mathbf{F}}{\arg\min} \, \left\| \mathcal{G} - {[\![ \mathbf{B}, \mathbf{D}, \mathbf{F} ]\!]}_{R} \right\|_{F}^{2} + \eta \left\| \mathbf{I} - \mathbf{B} \cdot \mathbf{D}^{\modified{\mathrm{T}}} \right\|_{F}^{2},
	\label{eq:BDF}
\end{equation}where $\eta$ denotes the regularization coefficient chosen by the user. We concatenate tensor $\mathcal{G}$ and matrix $\eta \cdot \mathbf{I}$ along the third dimension into a new tensor ${\mathcal{G}}'\in {{\mathbb{C}}^{R\times R\times (4M+1)}}$: 
\vspace{-5pt} 
\begin{align}
	{{{\mathcal{G}}'}_{(:,:,w)}} \triangleq 
	\begin{cases} 
		{{\mathcal{G}}_{(:,:,w)}}, & \text{if } w \in \{1,2,3,\ldots,4M\}, \\ 
		\eta \cdot \mathbf{I}, & \text{if } w = 4M+1.
	\end{cases}
		\label{eq:daGpie}
\end{align}

Then \eqref{eq:BDF} can be formulated as an LS based optimization problem for the CPD of ${\mathcal{G}}'$:
\vspace{-5pt} 
\begin{align}
	\{\tilde{\mathbf{B}},\tilde{\mathbf{D}},\tilde{\mathbf{{F}}}'\} = 
	\underset{\mathbf{B},\mathbf{D},\mathbf{{F}}'}{\mathop{\arg \min }}\, 
	\left\| {\mathcal{G}}' - {{\left[\!\left[ \mathbf{B},\mathbf{D},\mathbf{{F}}' \right]\!\right]}_{R}} \right\|_{F}^{2},
	\label{eq:BDFpie}
\end{align}where $\mathbf{{F}'}\in {{\mathbb{C}}^{(4M+1)\times R}}$ is the concatenation of $\mathbf{F}$ and $\eta \cdot {{\mathbf{1}}_{1\times R}}$ along the row dimension, with ${{\mathbf{1}}_{1\times R}}$ being an all-one row vector of length $R$.

 Since the J-EVD problem \eqref{eq:daGpie} is analogous to an overdetermined CPD problem, it can be addressed using the GEVD algorithm to derive an algebraic solution, as discussed in \textit{Remark 2}. By solving the J-EVD problem for all $m$, $m=1,\dots,M$, via the CPD of ${\mathcal{G}}'$, we obtain estimates of the factor matrix $\tilde{\mathbf{B}}$ and the generators of $\mathbf{A}_{x,1}^{(m)}, \mathbf{A}_{x,2}^{(m)}, \mathbf{A}_{y,1}^{(m)},$ and $\mathbf{A}_{y,2}^{(m)}$, $m=1,\dots,M$, located in the first through the $4M$-th rows of $\tilde{\mathbf{{F}}}'$. 
 
 After computing $\tilde{\mathbf{B}}$, we define $\mathbf{\Omega }_{r}^{(m)} \triangleq \operatorname{unvec}\big((\mathbf{T}_{(2)}^{(m)} \tilde{\mathbf{B}}^{-\modified{\mathrm{T}}})_{(:,r)}\big)\in \mathbb{C}^{I^{(m)} \times K}$ to estimate the factor matrices $\tilde{\mathbf{A}}^{(m)}$ and $\tilde{\mathbf{C}}^{(m)}$. The matrix $\mathbf{\Omega }_{r}^{(m)}$ is then approximated as a rank-1 matrix with $\mathbf{\Omega }_{r}^{(m)} = \tilde{\mathbf{a}}_{r}^{(m)} \tilde{\mathbf{c}}_{r}^{(m){\modified{\mathrm{T}}}}$, where $\tilde{\mathbf{a}}_{r}^{(m)}$ and $\tilde{\mathbf{c}}_{r}^{(m)}$ represent the dominant left and right singular vectors, respectively. Some considerations for computing J-EVD are summarized in the following remark.
 
\textit{Remark 2}: Given that the least squares (LS) based optimization problem for the CPD of ${\mathcal{G}}'$ has dimensions of $R\times R\times (4M+1)$ and is overdetermined, the J-EVD problem \eqref{eq:daGpie} can be solved algebraically using the GEVD. We implement both the GEVD and simultaneous generalized schur decomposition (SGSD) algorithms using functions from the Tensorlab 3.0 toolbox \cite{Tensorlab3.0}. The output from the GEVD algorithm serves as the initial value for the SGSD algorithm, which improves the accuracy of the factor matrix $\tilde{\mathbf{B}}$ estimate by providing a more precise semi-algebraic solution. 

\textit{Remark 3}: Regarding the choice of the regularization weight coefficient $\eta$, this parameter balances two factors: (i) the transformation that best fits the data ${\mathcal{G}}'$ in the LS sense, and (ii) the constraint $\mathbf{B} \cdot {{\mathbf{D}}^{\modified{\mathrm{T}}}} = \mathbf{I}$ required by J-EVD. In practice, to emphasize the first factor, $\eta$ is typically set to a small value, specifically $\eta = 1$ in our experimental settings, which ensures the exactness of the constraint $\modified{\mathbf{B}} \cdot {{\mathbf{D}}^{\modified{\mathrm{T}}}} = \mathbf{I}$.

\textit{Remark 4}: The computational complexity of the semi-algebraic J-EVD-based C-CPD algorithm for MS CPLsA mainly involves dimensionality reduction, target matrix construction, solving the J-EVD problem via GEVD, and computing the factor matrices $\tilde{\mathbf{A}}^{(m)}$ and $\tilde{\mathbf{C}}^{(m)}$. Dimensionality reduction via truncated SVD has a complexity of $\mathcal{O}(2(\sum_{m=1}^{M} I^{(m)} K T^2) + 11 T^3)$ flops. Target matrix construction involves pseudo-inversion and matrix multiplication, with a complexity of $\mathcal{O}(3(I' - 1) K R^2)$ flops per matrix, repeated for $4M$ matrices, yielding a total complexity of $\mathcal{O}(12M(I' - 1) K R^2)$ flops, where $I'$ is the maximum of $\{I_{x,1}^{(m)}, I_{x,2}^{(m)}, I_{y,1}^{(m)}, I_{y,2}^{(m)}, m=1, \dots, M\}$. Computing ${\mathbf{\Omega }}^{(m)} \triangleq \mathbf{T}_{(2)}^{(m)} \tilde{\mathbf{B}}^{-\modified{\mathrm{T}}}$ requires $\mathcal{O}(I^{(m)} K R (2R - 1))$ flops, repeated for $M$ matrices, with a total complexity of $\mathcal{O}(\sum_{m=1}^{M} I^{(m)} K R (2R - 1))$. The matrix $\mathbf{\Omega}_r^{(m)} \triangleq \operatorname{unvec}(\mathbf{\Omega}^{(m)}(:, r))$ is reshaped, and performing SVD on $\mathbf{\Omega}_r^{(m)}$ requires $\mathcal{O}(2K (I^{(m)})^2 + 11 (I^{(m)})^3)$ flops, performed for all $\mathbf{\Omega}_r^{(m)}$ matrices, leading to a total complexity of $\mathcal{O}(\sum_{m=1}^{M} R(2K (I^{(m)})^2 + 11 (I^{(m)})^3))$. Solving the J-EVD problem via GEVD requires $\mathcal{O}(30 R^2)$ flops for each pair of $\mathcal{G}(:,:,w_1)$ and $\mathcal{G}(:,:,w_2)$, where $w_1 \neq w_2$. The GEVD output initializes the SGSD algorithm, whose complexity depends on iteration count and convergence. \added{In contrast, the standard algebraic method for C-CPD based on simultaneous diagonalization exhibits significantly higher computational complexity. For the C-CPD model in \eqref{eq:Xm}, the core computational burden of the algebraic method proposed in \cite{Sorensen2015b}, which is based on simultaneous diagonalization, lies in the construction and manipulation of second-order compound matrices. Specifically, this method constructs the structured matrix from combinations of tensor slices, with a computational complexity of $\mathcal{O}(I^2 K^2 R^2)$ flops. Subsequently, it extracts a symmetric subspace basis via singular value decomposition, which incurs a complexity of $\mathcal{O}(I^2 K^2 R^4)$ flops. These operations dominate the total computational cost of the algebraic C-CPD method in \cite{Sorensen2015b}, and become particularly intensive when either the tensor dimensions or the rank $R$ are large.}

We summarize the proposed method in Algorithm \ref{alg:alg1}.

\begin{algorithm}[t]
	\caption{(Semi-) Algebraic C-CPD based on J-EVD for \added{MS} CPLsA}
	\parbox{\textwidth}{\textbf{Input}: (i) The C-CPD model $\{\mathcal{T}^{(m)}\}$ constructed from the MS CPLsA MIMO radar after dimensionality reduction; (ii) Number of targets: $R$.}
	\vspace{-0.82em} %
	\begin{algorithmic}
   \STATE \hspace{-0.55cm} \textbf{Step 1: Construct} $\mathbf{G}_{v}^{(m)}\!\in\! {{\mathbb{C}}^{R\times R}}$ \textbf{for fixed} $m$ \textbf{and} $v=1,\ldots,4$.
		\STATE (1) Check working conditions for this algorithm according to \textit{Remark 1}, \hfill \\ \hspace{0.52cm} \textit{Subsection}~\ref{subsec3.1}; 
		\STATE (2) Construct ${\mathbf{G}_{1}^{(m)}},\!{\mathbf{G}_{2}^{(m)}},\!{\mathbf{G}_{3}^{(m)}},$ and ${\mathbf{G}_{4}^{(m)}}$ by \eqref{eq:Txm}--\eqref{eq:G1G2G3G4}, \textit{Subsection}~\ref{subsec3.1}\modified{.} 
		\STATE  
\hspace{-0.55cm} \textbf{Step 2: Compute} $\tilde{\mathbf{B}}$ \textbf{via J-EVD of} $\{\mathbf{G}_{v}^{(m)},v=1,...,4,m=1,...,M\}$, \hfill \\ \hspace{-0.55cm}  \textbf{and} ${{\tilde{\mathbf{A}}}^{(m)}}$ \textbf{and} ${{\tilde{\mathbf{C}}}^{(m)}}$ \textbf{using rank-1 approximation}.
		\STATE \parbox{\textwidth}{(1) Perform J-EVD to ${\mathbf{G}_{1}^{(m)}},\!{\mathbf{G}_{2}^{(m)}},\!{\mathbf{G}_{3}^{(m)}}$ and ${\mathbf{G}_{4}^{(m)}}$ to obtain $\tilde{\mathbf{B}}$ via procedures \hfill } \\ \hspace{0.52cm} described in \eqref{eq:daG}--\eqref{eq:BDFpie}, \textit{Subsection}~\ref{subsec3.1}\modified{;}
		\STATE (2) Once $\tilde{\mathbf{B}}$ has been obtained, $\tilde{\mathbf{A}}^{(m)}$ and $\tilde{\mathbf{C}}^{(m)}$ are computed via rank-1\\ \hspace{0.52cm} approximation.
	\end{algorithmic}
	\textbf{Output}: Estimates of the factor matrices $\tilde{\mathbf{B}}$, $\tilde{\mathbf{A}}^{(m)}$ and $\tilde{\mathbf{C}}^{(m)}$, $m\!=\! 1,\ldots,M.$
	\label{alg:alg1}
\end{algorithm}

\subsection{(Semi-) Algebraic C-CPD based on J-EVD for \added{MS} CPPA}
\label{subsec3.2}
For the C-CPD of a third-order tensor in \eqref{eq:Xm}, we assume that the second factor matrix \(\mathbf{B}\) has dimensions \(R \times R\) via dimensionality reduction \cite{Gong2022}, \cite{Gong2018}, \cite{Gong2019}. In the proposed J-EVD-based C-CPD algorithm for \added{MS} CPPA, the Vandermonde structure in the first factor matrix, corresponding to the local sparse ULA configuration within CPPA, is utilized to construct target matrices for the J-EVD. The proposed algorithm consists of the following steps:

\noindent     	\textbf{Step 1: Construct} $\mathbf{G}_{v}^{(m)}\in {{\mathbb{C}}^{R\times R}}$ \textbf{for fixed} $m$ \textbf{and} $v=1,\ldots ,4$.  	

To begin, we define the following notations. The factor matrix $\mathbf{A}^{(m)}$, corresponding to the steering vector of the $m$-th receive array in the MS CPPA, is obtained via the Cartesian product of two sensor location sets along the $x$-axis and $y$-axis, and can be represented as $\mathbf{A}^{(m)} = \mathbf{A}_{x}^{(m)} \odot \mathbf{A}_{y}^{(m)}$, where $\mathbf{A}_{x}^{(m)}$ has dimensions $|\mathbb{S}_{x}^{(m)}| \times R$ and $\mathbf{A}_{y}^{(m)}$ is of $|\mathbb{S}_{y}^{(m)}| \times R$. For the two sparse ULAs along the $x$-axis, the Vandermonde matrices are defined as $\mathbf{A}_{x,1}^{(m)} = \mathbf{A}_{x}^{(m)}(\mathbb{Q}_{x,1}^{(m)},:)$ and $\mathbf{A}_{x,2}^{(m)} = \mathbf{A}_{x}^{(m)}(\mathbb{Q}_{x,2}^{(m)},:)$, where $\mathbb{Q}_{x,1}^{(m)}$ and $\mathbb{Q}_{x,2}^{(m)}$ are index sets within $\mathbb{S}_{x}^{(m)}$ corresponding to $\mathbb{S}_{x,1}^{(m)}$ and $\mathbb{S}_{x,2}^{(m)}$, respectively. Similarly, for the $y$-axis, the Vandermonde matrices are given by $\mathbf{A}_{y,1}^{(m)} = \mathbf{A}_{y}^{(m)}(\mathbb{Q}_{y,1}^{(m)},:)$ and $\mathbf{A}_{y,2}^{(m)} = \mathbf{A}_{y}^{(m)}(\mathbb{Q}_{y,2}^{(m)},:)$, where $\mathbb{Q}_{y,1}^{(m)}$ and $\mathbb{Q}_{y,2}^{(m)}$ represent the index sets within $\mathbb{S}_{y}^{(m)}$ corresponding to $\mathbb{S}_{y,1}^{(m)}$ and $\mathbb{S}_{y,2}^{(m)}$, respectively. Next, we construct $\mathbf{G}_{1}^{(m)}$ by leveraging the Vandermonde structure in $\mathbf{A}_{x,1}^{(m)}$. Given that $\mathbf{A}^{(m)} = \mathbf{A}_{x}^{(m)} \odot \mathbf{A}_{y}^{(m)}$, the C-CPD model in \eqref{eq:Xm} can be expressed as:
\vspace{-5pt} 
\begin{align}
	\mathcal{T}_{x}^{(m)} = {{[\![\mathbf{A}_{x}^{(m)},\mathbf{A}_{y}^{(m)},\mathbf{B},{{\mathbf{C}}^{(m)}}]\!]}_{R}}.
	\label{eq:Txm2}
\end{align}

Then, extract the data tensor corresponding to the first sparse ULA along the $x$-axis of the $m$-th receive array:
\vspace{-5pt} 
\begin{align}
	\mathcal{T}_{x,1}^{(m)} \triangleq \mathcal{T}_{x}^{(m)}(\mathbb{Q}_{x,1}^{(m)},:,:,:) = 
	{{[\![\mathbf{A}_{x,1}^{(m)},\mathbf{A}_{y}^{(m)},\mathbf{B},{{\mathbf{C}}^{(m)}}]\!]}_{R}}.
		\label{eq:Tx1m2}
\end{align}

After that, since $\mathcal{T}_{x,1}^{(m)}$ admits a fourth-order CPD, its mode-3 matrix representation $\mathbf{T}_{x,1}^{(m)}$ is:
\vspace{-5pt} 
\begin{align}
	\mathbf{T}_{x,1}^{(m)} = (\mathbf{A}_{x,1}^{(m)} \odot \mathbf{A}_{y}^{(m)} \odot {{\mathbf{C}}^{(m)}}) \cdot {{\mathbf{B}}^{\modified{\mathrm{T}}}}.
	\label{eq:TTx1m2}
\end{align}

We choose two submatrices $\mathbf{M}_{x,1}^{(m,1)}$ and $\mathbf{M}_{x,1}^{(m,2)}$ from $\mathbf{T}_{x,1}^{(m)}$, each having dimensions $(I_{x,1}^{(m)}-1){{I}_{y}}K\times R$:
\vspace{-5pt} 
\begin{align}
	\left\{ 
	\begin{aligned}
		 \mathbf{M}_{x,1}^{(m,1)} &\triangleq \mathbf{T}_{x,1}^{(m)}(1:(I_{x,1}^{(m)}-1){{I}_{y}}K,:) \\
		 \quad &= (\underline{\mathbf{A}}_{x,1}^{(m)} \odot \mathbf{A}_{y}^{(m)} \odot {\mathbf{C}^{(m)}}) \cdot {{\mathbf{B}}^{\modified{\mathrm{T}}}}, \\ 
		 \mathbf{M}_{x,1}^{(m,2)} &\triangleq \mathbf{T}_{x,1}^{(m)}({{I}_{y}}K+1:I_{x,1}^{(m)}{{I}_{y}}K,:) \\
		 \quad& = (\overline{\mathbf{A}}_{x,1}^{(m)} \odot \mathbf{A}_{y}^{(m)} \odot {\mathbf{C}^{(m)}}) \cdot {{\mathbf{B}}^{\modified{\mathrm{T}}}}.  
	\end{aligned}
	\right.
	\label{eq:Mx1m1Mx1m2}
\end{align}

Since $\mathbf{A}_{x,1}^{(m)}$ is a Vandermonde matrix, we have $\overline{\mathbf{A}}_{x,1}^{(m)}=\underline{\mathbf{A}}_{x,1}^{(m)}\cdot \mathbf{Z}_{x,1}^{(m)}$, where $\mathbf{Z}_{x,1}^{(m)}$ is a diagonal matrix holding the Vandermonde generators of $\mathbf{A}_{x,1}^{(m)}$, $z_{x1,1}^{(m)},...,z_{x1,R}^{(m)}$, in its main diagonal. Therefore, we have the following result:
\vspace{-5pt} 
\begin{align}
	\overline{\mathbf{A}}_{x,1}^{(m)} \odot \mathbf{A}_{y}^{(m)} \odot {\mathbf{C}^{(m)}} = 
	(\underline{\mathbf{A}}_{x,1}^{(m)} \odot \mathbf{A}_{y}^{(m)} \odot {\mathbf{C}^{(m)}}) \cdot \mathbf{Z}_{x,1}^{(m)}.
	\label{eq:Ax1Ay}
\end{align}

Assume that $\mathbf{M}_{x,1}^{(m,1)}$ has full column rank, and construct the target matrix as $\mathbf{G}_{1}^{(m)}\triangleq {{[{{(\mathbf{M}_{x,1}^{(m,1)})}^{\dagger }}\mathbf{M}_{x,1}^{(m,2)}]^{\modified{\mathrm{T}}}}}\in {{\mathbb{C}}^{R\times R}}$. Then, after simple derivations, we obtain:
\vspace{-5pt} 
\begin{align}
	\mathbf{G}_{1}^{(m)} \triangleq {{[ {{(\mathbf{M}_{x,1}^{(m,1)})}^{\dagger}} \mathbf{M}_{x,1}^{(m,2)}]^{\modified{\mathrm{T}}}}} 
	= \mathbf{B} \cdot \mathbf{Z}_{x,1}^{(m)} \cdot {{\mathbf{B}}^{-1}}.
	\label{eq:G1m2}
\end{align}

Next, we follow similar procedures to exploit the Vandermonde structure in $\mathbf{A}_{x,2}^{(m)}$, $\mathbf{A}_{y,1}^{(m)}$, and $\mathbf{A}_{y,2}^{(m)}$ to construct the matrices $\mathbf{G}_{2}^{(m)}$, $\mathbf{G}_{3}^{(m)}$, and $\mathbf{G}_{4}^{(m)}$, respectively. For instance, to construct $\mathbf{G}_{2}^{(m)}$, we replace $\mathbb{Q}_{x,1}^{(m)}$ in \eqref{eq:Tx1m2} with $\mathbb{Q}_{x,2}^{(m)}$ to obtain the data tensor $\mathcal{T}_{x,2}^{(m)}$ corresponding to the $x$-axis direction of the $m$-th receive array. Then, $\mathbf{T}_{x,2}^{(m)}$ is obtained by \eqref{eq:TTx1m2}. The submatrices of $\mathbf{T}_{x,2}^{(m)}$, namely $\mathbf{M}_{x,2}^{(m,1)}$ and $\mathbf{M}_{x,2}^{(m,2)}$, are constructed using \eqref{eq:Mx1m1Mx1m2}. Subsequently, $\mathbf{G}_{2}^{(m)}$ is computed with $\mathbf{M}_{x,2}^{(m,1)}$ and $\mathbf{M}_{x,2}^{(m,2)}$ using \eqref{eq:G1m2}. Next, we swap the first and second dimensions of $\mathcal{T}_{x}^{(m)}$ in \eqref{eq:Txm2} to construct a new tensor $\mathcal{T}_{y}^{(m)}$. Using the index sets $\mathbb{Q}_{y,1}^{(m)}$ and $\mathbb{Q}_{y,2}^{(m)}$, we extract the corresponding data tensors from \added{$\mathcal{T}_{y}^{(m)}$} and follow a similar process as described in \eqref{eq:Tx1m2}--\eqref{eq:G1m2} to construct $\mathbf{G}_{3}^{(m)}$ and $\mathbf{G}_{4}^{(m)}$. As a result, the matrices $\mathbf{G}_{1}^{(m)}, \mathbf{G}_{2}^{(m)}, \mathbf{G}_{3}^{(m)}$, and $\mathbf{G}_{4}^{(m)}$ can be collectively expressed in the following J-EVD formulation:
\vspace{-5pt} 
\begin{align}
	\left\{
	\begin{aligned}
		& \mathbf{G}_{1}^{(m)} = \mathbf{B} \cdot \mathbf{Z}_{x,1}^{(m)} \cdot {{\mathbf{B}}^{-1}}, \quad 
		\mathbf{G}_{2}^{(m)} = \mathbf{B} \cdot \mathbf{Z}_{x,2}^{(m)} \cdot {{\mathbf{B}}^{-1}}, \\ 
		& \mathbf{G}_{3}^{(m)} = \mathbf{B} \cdot \mathbf{Z}_{y,1}^{(m)} \cdot {{\mathbf{B}}^{-1}}, \quad 
		\mathbf{G}_{4}^{(m)} = \mathbf{B} \cdot \mathbf{Z}_{y,2}^{(m)} \cdot {{\mathbf{B}}^{-1}}, 
	\end{aligned}
	\right.
	\label{eq:G1G2G3G42}
\end{align}
where $\mathbf{Z}_{x,1}^{(m)},\mathbf{Z}_{x,2}^{(m)},\mathbf{Z}_{y,1}^{(m)},$ and $\mathbf{Z}_{y,2}^{(m)}$ are diagonal, holding in their main diagonals the Vandermonde generators of $\mathbf{A}_{x,1}^{(m)}, \mathbf{A}_{x,2}^{(m)}, \mathbf{A}_{y,1}^{(m)},$ and $\mathbf{A}_{y,2}^{(m)}$, respectively. Note that in this study, J-EVD aims to find an invertible matrix $\mathbf{B}$, such that ${{\mathbf{B}}^{-1}}\cdot \mathbf{G}_{v}^{(m)}\cdot \mathbf{B}$ is diagonal for $v\in \{1,2,3,4\}$.

\textit{Remark 5}: Note that at least one of matrices $\mathbf{M}_{x,1}^{(m,1)},\mathbf{M}_{x,2}^{(m,1)},\mathbf{M}_{y,1}^{(m,1)}$ and $\mathbf{M}_{y,2}^{(m,1)}$ for $m=1,…,M$, is required to have full column rank. This ensures that at least one target matrix is available for J-EVD (if only one target matrix is successfully constructed, J-EVD reduces to EVD). Generically, this is equal to requiring the following working conditions for the proposed J-EVD based C-CPD algorithm:
\vspace{-5pt} 
\begin{align}
	\left\{
	\begin{aligned}
		 \min (T, J) \geq & R, \\ 
		\modified{{I}''K} \geq & R,  
	\end{aligned}
	\right.
	\label{eq:condition2}
\end{align}
where ${I}''$ represents the value corresponding to the maximum element in the set $\{(I_{x,1}^{(m)}-1)I_{y}^{(m)},(I_{x,2}^{(m)}-1)I_{y}^{(m)},$ $(I_{y,1}^{(m)}-1)I_{x}^{(m)},(I_{y,2}^{(m)}-1)I_{x}^{(m)},m=1,...,M\}$. These conditions can be regarded as generic sufficient uniqueness conditions for the C-CPD problem in \eqref{eq:Xm} of the proposed J-EVD based method for the MS CPPA based MIMO radar structure.

\noindent     	\textbf{Step 2: Compute} $\mathbf{B}$ \textbf{via J-EVD of} $\{\mathbf{G}_{v}^{(m)},v=1,...,4,m=1,...,M\}$, \textbf{and} ${{\mathbf{A}}^{(m)}}$ \textbf{and} ${{\mathbf{C}}^{(m)}}$ \textbf{using rank-1 approximation}.

Due to the main difference between MS CPLsA and MS CPPA being in their sensor locations, only Step 1 differs in the proposed J-EVD-based semi-algebraic C-CPD algorithm, while Step 2 remains the same. Therefore, for Step 2 of the MS CPPA algorithm, the computations follow \eqref{eq:daG}--\eqref{eq:BDFpie}, as described in \textit{Subsection} \ref{subsec3.1}. As Step 2 is identical for both configurations, some considerations for computing J-EVD can be found in \textit{Remarks 2} and \textit{3} in \textit{Subsection} \ref{subsec3.1}. Additionally, \textit{Remark 6} provides specific details on the computational complexity of the proposed algorithm for MS CPPA.

Once $\tilde{\mathbf{B}}$ is obtained, $\tilde{\mathbf{A}}^{(m)}$ and $\tilde{\mathbf{C}}^{(m)}$ are computed via rank-1 approximation, as detailed in Step 2 of \textit{Subsection} \ref{subsec3.1}.

\textit{Remark 6}: The computational complexity of the proposed \added{MS} CPPA algorithm primarily involves dimensionality reduction, target matrix construction, solving the J-EVD problem via GEVD, and computing the factor matrices \(\tilde{\mathbf{A}}^{(m)}\) and \(\modified{\tilde{\mathbf{C}}^{(m)}}\). \modified{The only difference between MS CPPA and MS CPLsA, as noted in \textit{Remark 4}, lies in the target matrix construction, which requires approximately \(\mathcal{O}(I'' I_y K R^2)\) flops per matrix.} For \(4M\) matrices, the overall complexity is \(\mathcal{O}(12M I'' K R^2)\) flops, where \(I''\) is the maximum of \(\{(I_{x,1}^{(m)}-1)I_y^{(m)}, (I_{x,2}^{(m)}-1)I_y^{(m)}, (I_{y,1}^{(m)}-1)I_x^{(m)}, (I_{y,2}^{(m)}-1)I_x^{(m)},\, m=1,\dots, M\}\).  \added{In comparison, the standard algebraic C-CPD algorithm based on simultaneous diagonalization, such as the one proposed in \cite{Sorensen2015b}, involves constructing second-order compound matrices and extracting symmetric subspace bases via SVD, resulting in a significantly higher complexity of up to \(\mathcal{O}(I^2 K^2 R^4)\) flops, as detailed in \textit{Remark 4}. This highlights the computational advantage of the proposed semi-algebraic framework in scenarios with large array dimensions or high model rank.}

We summarize the proposed method in Algorithm \ref{alg:alg2}.

\begin{algorithm}[h]
	\caption{(Semi-) Algebraic C-CPD based on J-EVD for \added{MS} CPPA}
	\parbox{\textwidth}{\textbf{Input}: (i) The C-CPD model \( \{\mathcal{T}^{(m)}\} \) constructed from the MS CPPA MIMO radar after dimensionality reduction; (ii) Number of targets: $R$.}
	\vspace{-0.82em} %
	\begin{algorithmic}
		\STATE
		\hspace{-0.55cm} \textbf{Step 1: Construct} $\mathbf{G}_{v}^{(m)}\!\in\! {{\mathbb{C}}^{R\times R}}$ \textbf{for fixed} $m$ \textbf{and} $v=1,\ldots,4$.
		\STATE (1) Check working conditions for this algorithm according to \textit{Remark 5}, \hfill \\ \hspace{0.52cm} \textit{Subsection}~\ref{subsec3.2}; 
		\STATE (2) Construct ${\!\mathbf{G}_{1}^{(m)}},\!{\mathbf{G}_{2}^{(m)}},\!{\mathbf{G}_{3}^{(m)}},\!$ and $\!{\mathbf{G}_{4}^{(m)}\!}$ by \eqref{eq:Txm2}--\eqref{eq:G1G2G3G42}, \textit{Subsection}~\ref{subsec3.2}\modified{.} 
		\STATE  
		\hspace{-0.55cm} \textbf{Step 2: Compute} $\tilde{\mathbf{B}}$ \textbf{via J-EVD of} $\{\mathbf{G}_{v}^{(m)},v=1,...,4,m=1,...,M\}$, \hfill \\ \hspace{-0.55cm}  \textbf{and} ${{\tilde{\mathbf{A}}}^{(m)}}$ \textbf{and} ${{\tilde{\mathbf{C}}}^{(m)}}$ \textbf{using rank-1 approximation}.
        \STATE{(1) Perform J-EVD on ${\!\mathbf{G}_{1}^{(m)}}\!,\!{\mathbf{G}_{2}^{(m)}}\!,\!{\mathbf{G}_{3}^{(m)}}\!\!$ and $\!{\mathbf{G}_{4}^{(m)}\!\!}$ to obtain $\tilde{\!\mathbf{B}}\!$ using the pro-}  \\ \hspace{0.52cm}  {cedures described in \eqref{eq:daG}--\eqref{eq:BDFpie}, \textit{Subsection}~\ref{subsec3.1}. This step follows the} \\ \hspace{0.52cm} {same procedure as in MS CPLsA, since the operations are identical\modified{;}}
		\STATE (2) Once $\tilde{\mathbf{B}}$ has been obtained, $\tilde{\mathbf{A}}^{(m)}$ and $\tilde{\mathbf{C}}^{(m)}$ are computed via rank-1\\ \hspace{0.52cm} approximation.
	\end{algorithmic}
	\textbf{Output}: Estimates of the factor matrices $\tilde{\mathbf{B}}$, $\tilde{\mathbf{A}}^{(m)}$ and $\tilde{\mathbf{C}}^{(m)}$, $m\!=\! 1,\ldots,M.$
	\label{alg:alg2}
\end{algorithm}

\subsection{Optimization based C-CPD}
Besides the (semi-)algebraic J-EVD based C-CPD algorithm, as proposed in \modified{Subsections \ref{subsec3.1} and \ref{subsec3.2} for MS CPLsA and MS CPPA}, respectively, optimization-based C-CPD algorithms also exist. In addition, alternating least squares (ALS) or nonlinear least squares (NLS) iterations can be employed to compute the C-CPD. For further details on these methods, please refer to \cite{Gong2018,Gong2019,Sorensen2015a,Sorensen2015b,Sorber2015}.

Comparing (semi-)algebraic and optimization-based C-CPD algorithms, the former guarantees exact solutions under ideal conditions but may perform suboptimally in noisy environments. The latter aims to minimize fitting error and reach the global minimum but can be sensitive to initialization, especially in underdetermined scenarios, and may not always provide accurate results, even in noiseless conditions. Therefore, using the algebraic algorithm for efficient initialization in optimization-based methods is recommended for practical applications.

\subsection{Target Localization}

To determine the target locations, we use the estimated factor matrices (i.e., the receive steering vectors) in a two-step process:

\text{(i) DOA estimation:} For MS CPLsA, after applying the (semi-)algebraic C-CPD algorithm based on J-EVD, the factor matrices \(\tilde{\mathbf{A}}^{(m)}\) for all receive arrays, \(m = 1, \dots, M\), are obtained. The steering vectors along the \(x\)- and \(y\)-axes for the \(m\)-th array are extracted as: $\tilde{\mathbf{A}}_{x}^{(m)} = \tilde{\mathbf{A}}^{(m)}(\mathbb{Q}_{x}^{(m)}, :)$ and $\tilde{\mathbf{A}}_{y}^{(m)}=\tilde{\mathbf{A}}^{(m)} (\mathbb{Q}_{y}^{(m)}, :),$ where $\mathbb{Q}_{x}^{(m)}$ and $\mathbb{Q}_{y}^{(m)}$ are index sets corresponding to the \(x\) and \(y\) directions, respectively. These steering vectors form a virtual CPPA steering matrix: $\hat{\mathbf{A}}^{(m)} = \tilde{\mathbf{A}}_{x}^{(m)} \odot \tilde{\mathbf{A}}_{y}^{(m)}.$ Then, the DOA parameters for each target are computed from \(\hat{\mathbf{A}}^{(m)}\) using the single-source MHR method \cite{Sorensen2016, Sorensen2018}.

For MS CPPA, unlike CPLsA, it is not necessary to compute the virtual CPPA steering matrix. Instead, after obtaining the estimated factor matrix \(\tilde{\mathbf{A}}^{(m)}\), we directly apply the single-source MHR method \cite{Sorensen2016, Sorensen2018} to \(\tilde{\mathbf{A}}^{(m)}\) to estimate the DOA parameters of each target.

\text{(ii) Target position estimation:} For each target, we fuse the DOA estimates from the various receive arrays to determine the target's position, as described in \cite{Gong2022}. Specifically, let \(\boldsymbol{\xi} \triangleq [\xi_x, \xi_y, \xi_z]^{\modified{\mathrm{T}}}\) be a point in space, and let \(\mathbf{p}_r^{(m)} \in \mathbb{R}^3\) denote the location of the center of the \(m\)-th receive array. The squared distance between \(\boldsymbol{\xi}\) and the line passing through \(\mathbf{p}_r^{(m)}\) in the direction of the DOA \(\mathbf{v}_{(\tilde{\theta}_r^{(m)}, \tilde{\varphi}_r^{(m)})}\) is given by
$
d_r^{(m)2} = || \boldsymbol{\xi} - \mathbf{p}_r^{(m)} ||^2 - || (\boldsymbol{\xi} - \mathbf{p}_r^{(m)})^{\modified{\mathrm{T}}} \mathbf{v}_{(\tilde{\theta}_r^{(m)}, \tilde{\varphi}_r^{(m)})} ||^2.
$ The target location is the point that minimizes the sum of these squared distances, i.e., the point closest to all estimated lines defined by the receive array positions and associated DOAs:
\vspace{-5pt} 
\begin{equation}
	\tilde{\boldsymbol{\xi}}_r = \arg \min_{\boldsymbol{\xi}} \sum_{m=1}^M \left( d_r^{(m)} \right)^2.
\end{equation}

We summarize the proposed method in Algorithm \ref{alg3}.

\begin{algorithm}[t]
	\caption{(Semi-)Algebraic J-EVD Based C-CPD with MS Coprime Arrays for Target Localization}
	\parbox{\textwidth}{\textbf{Input}:(i) Output of an MS MIMO radar with coprime arrays for each pulse period in one CPI:$\{\mathbf{X}_{k}^{(m)}\!,\! k \!=\! 1\!,\ldots\!,\!K,\! m \!=\! 1,\!\ldots\!,\! M\}$; (ii) Number of targets: $R$.}
	\vspace{-0.82em} %
	\begin{algorithmic}
		\STATE
		\hspace{-0.55cm} \textbf{1. Dimensionality reduction}.		
		\STATE (1) Construct third-order tensor C-CPD \eqref{eq:Xm} via stacking matrices $\mathbf{X}_{k}^{(m)}$;
		\STATE (2) Perform dimensionality reduction for the C-CPD model. 
		\STATE
		\hspace{-0.55cm} \textbf{2. Apply (semi-)algebraic C-CPD based on J-EVD for coprime arrays to obtain initial estimates.} 		
		\STATE \quad - For \added{MS} CPLsA, see Algorithm \ref{alg:alg1}. 
		\STATE \quad - For \added{MS} CPPA, see Algorithm \ref{alg:alg2}.
    	\STATE
		\hspace{-0.35cm}\textbf{3. Apply optimization-based C-CPD initialized with results of (semi-)algebraic J-EVD based C-CPD.}  
		\STATE (1) Initialize ALS/NLS optimization by the results of J-EVD based (semi-\hfill \\ \hspace{0.54cm} )algebraic C-CPD;
		\STATE (2) Iteratively update the factor matrices using an unconstrained ALS/ \hfill \\ \hspace{0.54cm} NLS approach until convergence.
		\STATE
		\hspace{-0.35cm}\textbf{4. Localize each target using estimates of factor matrices  $\tilde{\mathbf{A}}^{(m)}.$} 
		\STATE (1) Compute DOA estimates using the MHR method \cite{Sorensen2016, Sorensen2018};  
		\STATE \quad - For \added{MS} CPLsA, use matrix \(\hat{\mathbf{A}}^{(m)} = \tilde{\mathbf{A}}_{x}^{(m)} \odot \tilde{\mathbf{A}}_{y}^{(m)}\) from \(\tilde{\mathbf{A}}^{(m)}\). 
		\STATE \quad - For \added{MS} CPPA, apply directly to estimates of the factor matrices \(\!\tilde{\mathbf{A}}^{(m)}\!\!\).    
		\STATE (2) Determine target locations by fusing DOAs as described in \cite{Gong2022}.
	\end{algorithmic}
	\textbf{Output}: Estimates of target locations ${\tilde{\boldsymbol{\xi}}_{1},...,\tilde{\boldsymbol{\xi}}_{R}}$.
	\label{alg3}
\end{algorithm}

\section{Simulations}
\label{sec4}

\modified{This section evaluates the performance of the proposed target localization method and compares it with existing tensor-based approaches under} \modified{both overdetermined and underdetermined scenarios.} Specifically, we propose J-EVD-based semi-algebraic C-CPD algorithms for two distinct MS array configurations: MS CPLsA and MS CPPA. The comparative experiments for each configuration are presented in \textit{Subsections}~\ref{sec:CaseA} and~\ref{sec:CaseB}, respectively. The experiments use an MS coprime array MIMO radar system with one transmit array and three receive arrays, where \( I = I^{(1)} = I^{(2)} = I^{(3)} \). The settings are presented in Table~\ref{Settings}. All experiments satisfy \( \min(T,J) \geq R \) and are classified as: (1) overdetermined, where \( \min(I,J) \geq R \); or (2) single underdetermined, where \( \min(I,J) < R \) and \( \max(I,J) \geq R \). Notably, since this study focuses on scenarios with fast-moving targets, all experiments are configured with \( K \leq R \).

\begin{table}[htbp]
	\centering
	\caption{\textsc{Experiment Settings}}
	\label{Settings}
	\begin{tabular}{c ccccc c} 
		\toprule
		\multirow{2}{*}{Index} 
		& \multicolumn{5}{c}{Parameters} 
		& \multirow{2}{*}{\added{Scenario}} \\
		\cmidrule(lr){2-6}
		& \textit{I} & \textit{J} & \textit{K} & \textit{R} & \textit{T} &  \\
		\midrule
		A-1 & 13 & 27 & 8  & 10  & 64 & Overdetermined \\
		A-2 & 13 & 27 & 10 & 15  & 64 & Slightly single underdetermined \\
		A-3 & 13 & 27 & 25 & 25  & 64 & Highly single underdetermined \\
		B-1 & 25 & 49 & 5  & 20  & 64 & Overdetermined \\
		B-2 & 25 & 49 & 15 & 30  & 64 & Slightly single underdetermined \\
		B-3 & 25 & 49 & 45 & 45  & 64 & Highly single underdetermined \\
		\bottomrule
	\end{tabular}
\end{table}

In the experiments, we compare the proposed method with existing tensor-based algorithms C-CPD and CPD. All methods employ identical tensor construction and target localization steps, differing only in their tensor decomposition approaches. The methods compared are:
\begin{itemize}
	\renewcommand{\labelitemi}{\textbullet} 
 \vspace{-8pt}	
	\item C-CPD-JEVD: Computes the C-CPD model for the tensor set $\{{\mathcal{T}}^{(m)}, m \allowbreak = 1, \dots, M\}$ using the proposed semi-algebraic C-CPD algorithm based on J-EVD.
 \vspace{-8pt} 
	\item C-CPD-ALS(ALG): An ALS-based C-CPD method implemented via the \texttt{ccpd\_als.m} function with default settings in Tensorlab$^+$ \cite{Tensorlab2022}, initialized with results from C-CPD-JEVD.
 \vspace{-8pt} 
	\item C-CPD-MHR-SD: A C-CPD model for single-transmit, multi-receive MS MIMO radar, based on the semi-algebraic method proposed in \cite{Sorensen2017b}.
 \vspace{-22pt} 
	\item C-CPD-ALS(Rand): An ALS-based C-CPD method implemented via the \texttt{ccpd\_als.m} function with default settings in Tensorlab$^+$ \cite{Tensorlab2022}, initialized with random values.
 \vspace{-8pt} 
	\item CPD-ALS(ALG): An ALS-based CPD method implemented using the \texttt{cpd\_als.m} function with default settings in Tensorlab 3.0 \cite{Tensorlab3.0}, initialized with results from the algebraic CPD algorithm \cite{DeLathauwer2006}, via the \texttt{cpd3\_sd.m} function with default settings in Tensorlab 3.0 \cite{Tensorlab3.0}.
\vspace{-8pt} 
\end{itemize}

The output data of each receive array of an ST-MR-MP MIMO radar is then constructed as follows:
\vspace{-5pt} 
\begin{equation}
	{{\mathcal{Y}}^{(m)}}={{\sigma }_{s}}\frac{{{\mathcal{X}}^{(m)}}}{{{\left\| {{\mathcal{X}}^{(m)}} \right\|}_{F}}}+{{\sigma }_{n}}\frac{{{\mathcal{N}}^{(m)}}}{{{\left\| {{\mathcal{N}}^{(m)}} \right\|}_{F}}}, \quad m=1,\ldots ,M.
\end{equation}
Here \( {\mathcal{X}}^{(m)} \) denote the signal part of the observed data, where the $k$-th frontal slice, $\mathbf{X}_{k}^{(m)}$, is generated according to \eqref{eq:Xm}, and \( {\mathcal{N}}^{(m)} \) represents the noise tensor, with each entry drawn from a complex Gaussian distribution with zero mean and unit variance. The parameters \( {\sigma}_{s} \) and \( {\sigma}_{n} \) represent the signal and noise levels, respectively. In the generation of \( \mathbf{X}_{k}^{(m)} \), as described in \eqref{eq:Xm}, both the real and imaginary parts of each entry in the probing signal matrix \( \mathbf{S} \) are independently drawn from a Gaussian distribution with zero mean and unit variance. The RCS coefficients \( c_{k,r}^{(m)} \) are randomly generated from a complex Gaussian distribution with zero mean and unit variance, ensuring that the overall multi-pulse model adheres to the Swerling II model.

For all simulation experiments, whether using MS CPLsA or MS CPPA configurations, the MIMO radar consists of one transmit array and three receive arrays. The transmit array is located at \( (0, -8000\lambda, 0) \), and the receive arrays are at \( (-8000\lambda, 8000\lambda, 0) \), \( (0, 8000\lambda, 0) \), and \( (8000\lambda, 8000\lambda, 0) \), respectively. Targets are randomly distributed within a region where the \( x \)-coordinate ranges from \( -7000\lambda \) to \( 7000\lambda \), the \( y \)-coordinate from \( -7000\lambda \) to \( 7000\lambda \), and the \( z \)-coordinate from \( 4000\lambda \) to \( 8000\lambda \).

The signal-to-noise ratio (SNR) is defined as follows:
\vspace{-3pt} 
\begin{equation}
	SNR \triangleq 20 \lg (\sigma_s/\sigma_n).
\end{equation}

The performance of the compared methods is evaluated in terms of the mean angular error (MAE) and average \modified{CPU} time, where the MAE is defined as follows \cite{Gong2022}:
\vspace{-5pt} 
\begin{equation}
	{{\varepsilon }_{MAE}} \triangleq \frac{1}{MR} ( \sum_{m=1}^{M} \sum_{r=1}^{R} \arccos (| \mathbf{v}_{(\theta _{r}^{(m)}, \varphi _{r}^{(m)})}^{\modified{\mathrm{T}}} \mathbf{v}_{(\tilde{\theta }_{r}^{(m)}, \tilde{\varphi }_{r}^{(m)})}|)).
		\label{MAE}
\end{equation} 	
Here ${{\mathbf{v}}_{(\theta_{r}^{(m)},\varphi_{r}^{(m)})}}$ and ${{\mathbf{v}}_{({{\alpha }_{r}},{{\beta }_{r}})}}$ are the true DOA and DOD of the $r$-th target with respect to the $m$-th receive array and the $n$-th transmit array, respectively, and ${{\mathbf{v}}_{(\tilde{\theta }_{r}^{(m)},\tilde{\varphi }_{r}^{(m)})}}$ and ${{\mathbf{v}}_{({{{\tilde{\alpha }}}_{r}},{{{\tilde{\beta }}}_{r}})}}$ denote their estimates. 

Alongside the MAE curves of the compared methods, the Cramér-Rao bound (CRB)-based lower bounds for MAE, labeled as CRB-MAE and defined in \cite{Gong2022}, are also plotted in some relevant figures. The CRB-MAE is calculated based on \eqref{MAE}, where the estimates of $\theta_{r}^{(m)}$ and $\varphi_{r}^{(m)}$ are given by $\tilde{\theta}_{r,CRB}^{(m)} = \theta_{r}^{(m)} \pm \Delta \theta_{r}^{(m)}$ and $\tilde{\varphi}_{r,CRB}^{(m)} = \varphi_{r}^{(m)} \pm \Delta \varphi_{r}^{(m)}$, respectively. Here, $\Delta \theta_{r}^{(m)}$ and $\Delta \varphi_{r}^{(m)}$ represent the square roots of the CRB for $\theta_{r}^{(m)}$ and $\varphi_{r}^{(m)}$, respectively.

We note that each point in the MAE and CPU time curves represents the average of 200 Monte Carlo runs. All experiments are performed on a workstation with the following specifications: Intel (R) Core (TM) i7-12700H (12th Gen) CPU running at 2.30\,GHz, 40\,GB of RAM, 64-bit Windows 11 operating system, and M\textsc{atlab} version R2023a.

\subsection{Experiments for MS CPLsA}
\label{sec:CaseA}
In this subsection, we evaluate the target localization performance of the compared algorithms in an MS CPLsA MIMO radar system with the same array configuration. For the receive array, $M_x \!=\! M_y \!=\! 4, N_x \!=\! N_y \!=\! 7, I_{x,1}^{(m)} \!=\! I_{x,2}^{(m)} \!=\! I_{y,1}^{(m)} \!=\! I_{y,2}^{(m)} \!=\! 4$, and $I \!=\! 13$. For the transmit array, $M_x \!=\! M_y \!=\! 3, N_x \!=\! N_y \!=\! 5, J_{x,1} \!=\! J_{x,2} \!=\! J_{y,1} \!=\! J_{y,2}\! =\! 8$, and $J \!=\! 27$. Here, $J_{x,1}, J_{x,2}, J_{y,1}$, and $J_{y,2}$ refer to the number of elements in the sparse ULAs along the x- and y-axes of the transmit array, corresponding to the values in the \(m\)-th receive array.

We consider two types of scenarios: the overdetermined case and the single underdetermined case. Specifically, in \textit{Experiment A-1}, we set \( I = 13 \), \( J = 27 \), \( K = 8 \), and \( R = 10 \), corresponding to an \textit{overdetermined} case since \( \min(I, J) > R \). In \textit{Experiment A-2}, \( I = 13 \), \( J = 27 \), \( K = 10 \), and \( R = 15 \), representing a \textit{slightly single underdetermined} case where \( I < R \), and \( I \) is close to \( R \). In \textit{Experiment A-3}, \( I = 13 \), \( J = 27 \), \( K = 25 \), and \( R = 25 \), where \( J > R \), but \( R \text{ is close to } 2I \), thus defining this as a \textit{highly single underdetermined} case. In \textit{Experiment A-1}, the SNR varies from -25 dB to 5 dB; in \textit{Experiment A-2}, it varies from -20 dB to 10 dB; and in \textit{Experiment A-3}, it is from -10 dB to 20 dB.

\begin{figure}[t]
	\centering
	\begin{subfigure}[]{0.45\textwidth}
		\centering
		\includegraphics[width=\textwidth]{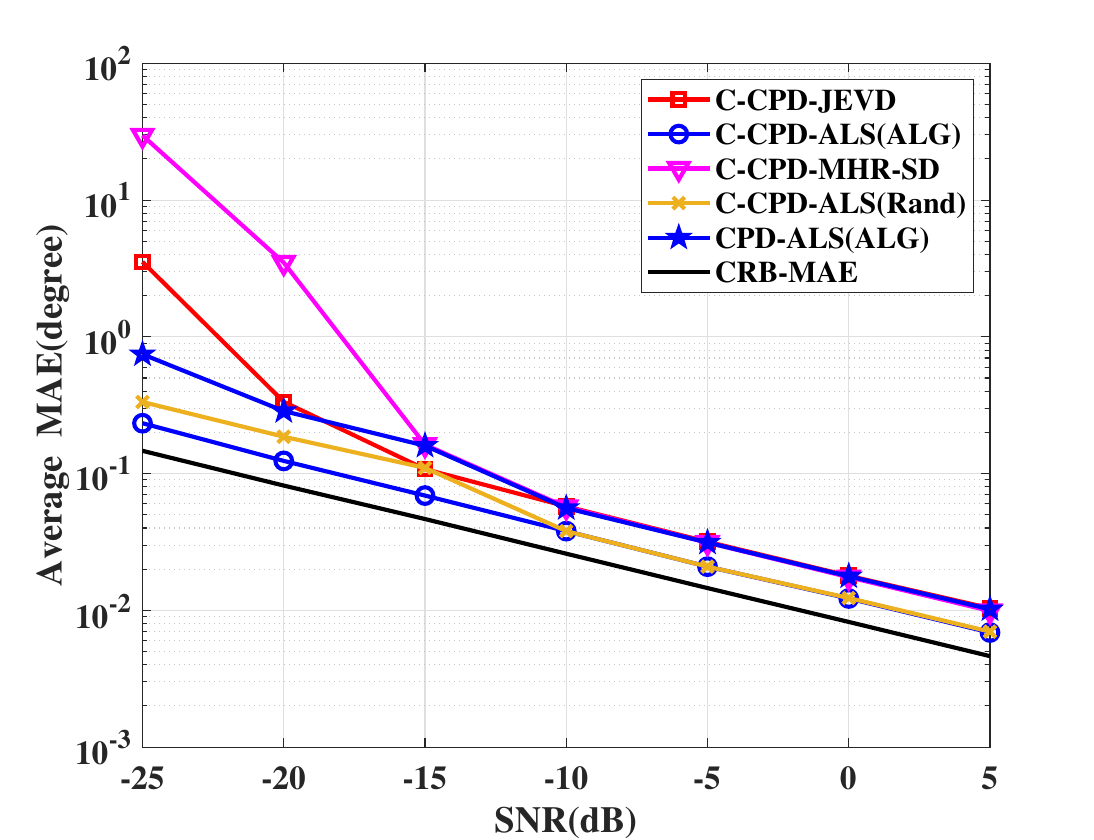}
		\caption{}
		\label{A_1_a}
	\end{subfigure}
	\begin{subfigure}[]{0.45\textwidth}
		\centering
		\includegraphics[width=\textwidth]{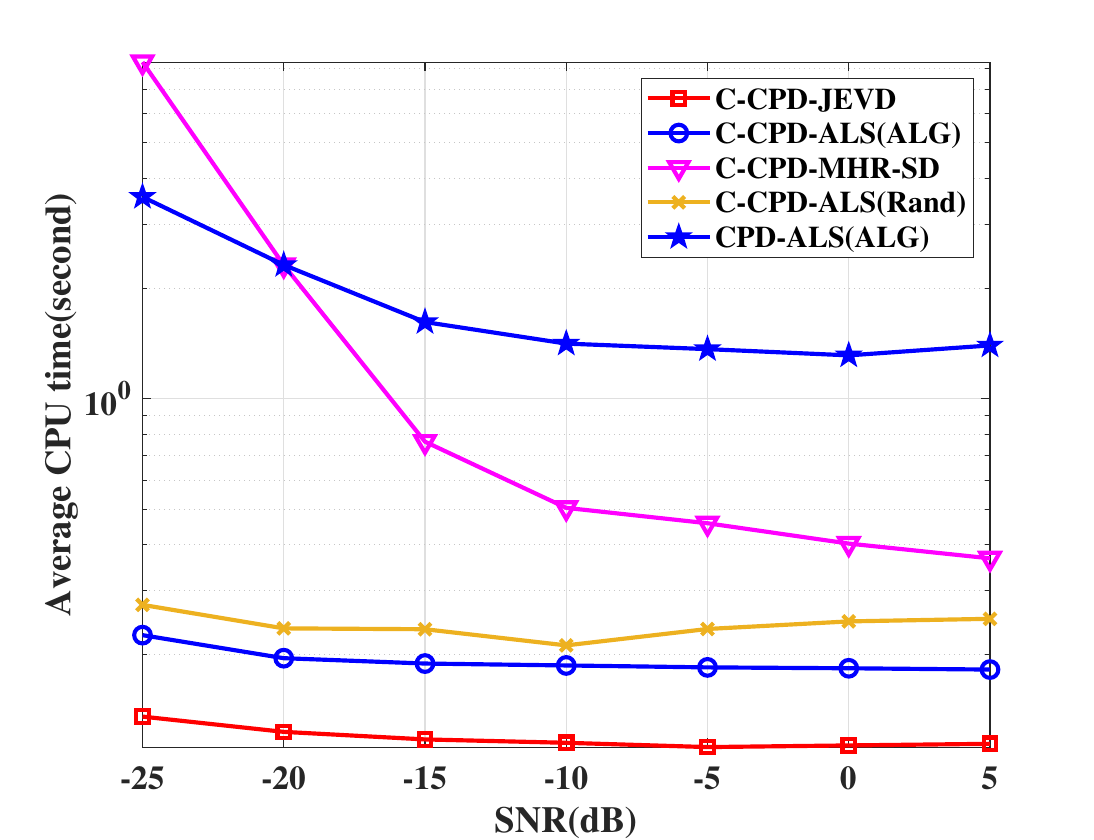}
		\caption{}
		\label{A_1_b}
	\end{subfigure}
	\caption{(a) Average MAE and (b) average CPU time vs. SNR in \textit{Experiment A-1}.}
	\label{A_1}
\end{figure}

Figs.~\ref{A_1_a} and~\ref{A_1_b} show the average MAE and CPU time versus SNR for \textit{Experiment A-1}. In this overdetermined scenario (Fig.~\ref{A_1}), the proposed semi-algebraic C-CPD-JEVD method outperforms C-CPD-MHR-SD, particularly at low SNRs (\(-25\) dB to \(-10\) dB). For tensor-based semi-algebraic methods, MAE performance is similar between \(-10\) dB and \(5\) dB due to the well-conditioned nature of the overdetermined scenario. Optimization-based C-CPD methods, such as C-CPD-ALS(ALG) and C-CPD-ALS(Rand), outperform semi-algebraic methods, with C-CPD-ALS(ALG) achieving the best MAE across all SNR ranges. In terms of CPU time, C-CPD-JEVD exhibits the lowest computational cost, and C-CPD-ALS(ALG), initialized with accurate values from C-CPD-JEVD, converges faster than C-CPD-ALS(Rand), which relies on random initialization.

\begin{figure}[t]
	\centering
	\begin{subfigure}[]{0.45\textwidth}
		\centering
		\includegraphics[width=\textwidth]{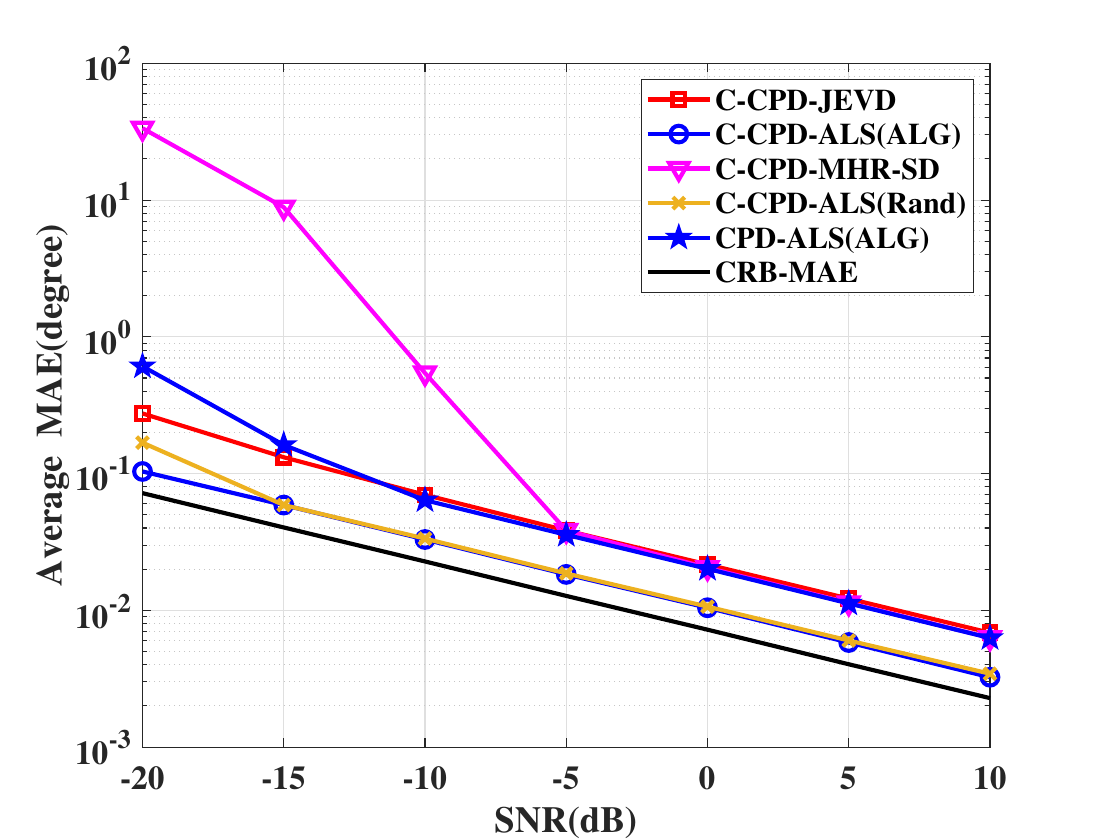}
		\caption{}
		\label{A_2_a}
	\end{subfigure}
	\begin{subfigure}[]{0.45\textwidth}
		\centering
		\includegraphics[width=\textwidth]{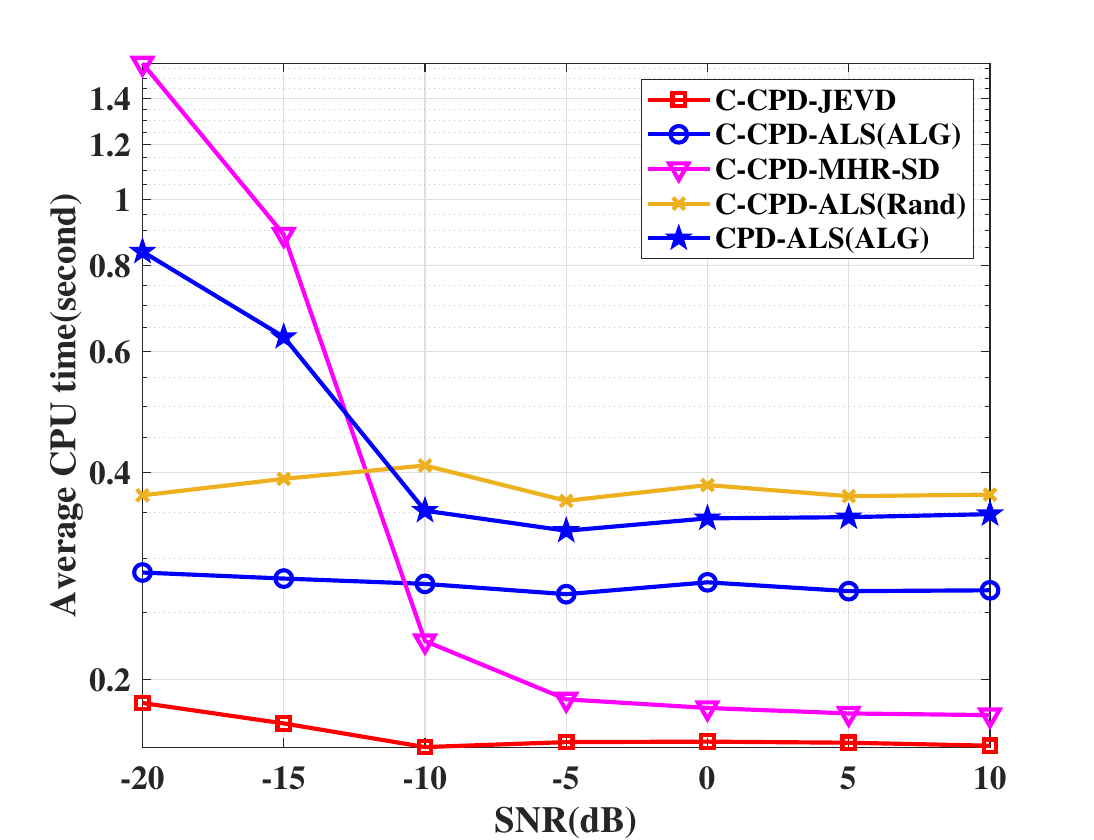}
		\caption{}
		\label{A_2_b}
	\end{subfigure}
	\caption{(a) Average MAE and (b) average CPU time vs. SNR in \textit{Experiment A-2}.}
	\label{A_2}
\end{figure}

Figs.~\ref{A_2_a} and~\ref{A_2_b} show the average MAE and CPU time versus SNR for the methods in \textit{Experiment A-2}. In the slightly single underdetermined case of Fig.~\ref{A_2}, C-CPD-ALS(ALG) performs best, followed by C-CPD-ALS(Rand), C-CPD-JEVD, CPD-ALS(ALG), and C-CPD-MHR-SD in terms of MAE. For semi-algebraic methods, the performance gap between C-CPD-JEVD and C-CPD-MHR-SD is larger in Fig.~\ref{A_2_a} than in Fig.~\ref{A_1_a}, especially at low SNR values (\(-20\) dB to \(-5\) dB). Among optimization-based approaches, C-CPD-ALS(ALG) outperforms C-CPD-ALS(Rand) and CPD-ALS(ALG), with faster convergence and better accuracy. In terms of CPU time, the results remain the same as those from \textit{Experiment A-1}, with C-CPD-JEVD maintaining the lowest computational cost.

\begin{figure}[t]
	\centering
	\begin{subfigure}[]{0.45\textwidth}
		\centering
		\includegraphics[width=\textwidth]{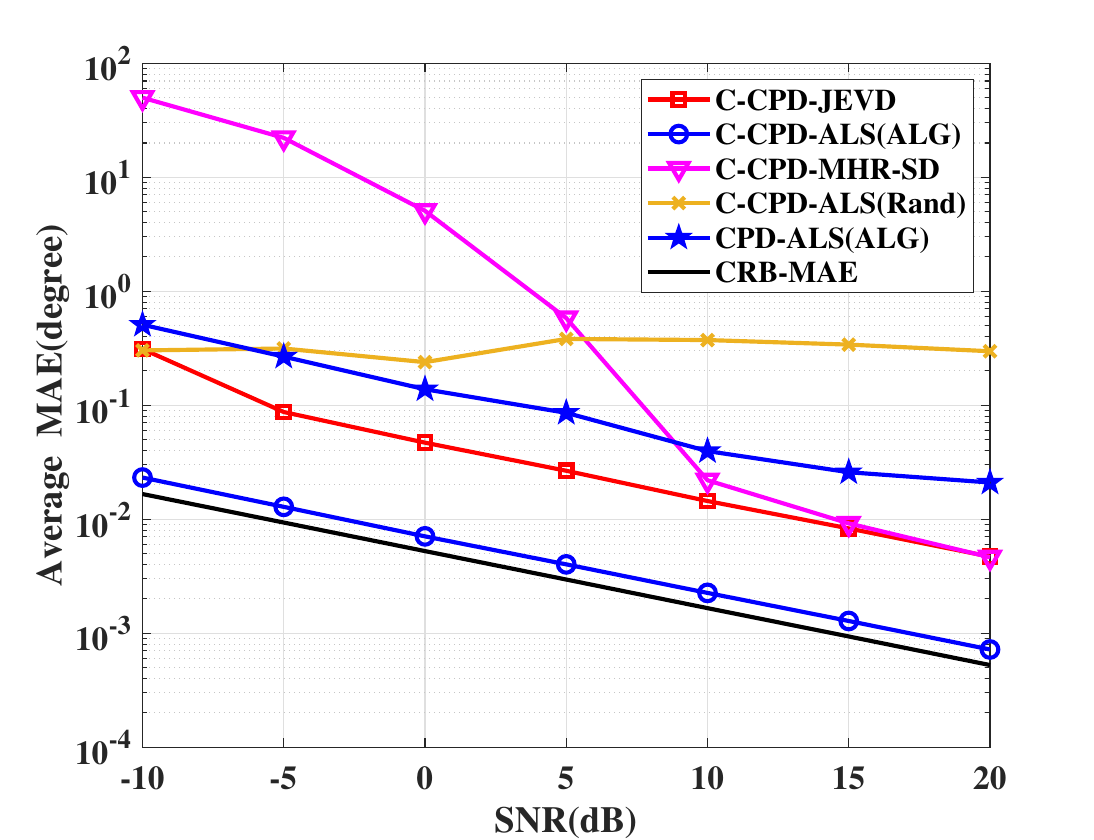}
		\caption{}
		\label{A_3_a}
	\end{subfigure}
	\begin{subfigure}[]{0.45\textwidth}
		\centering
		\includegraphics[width=\textwidth]{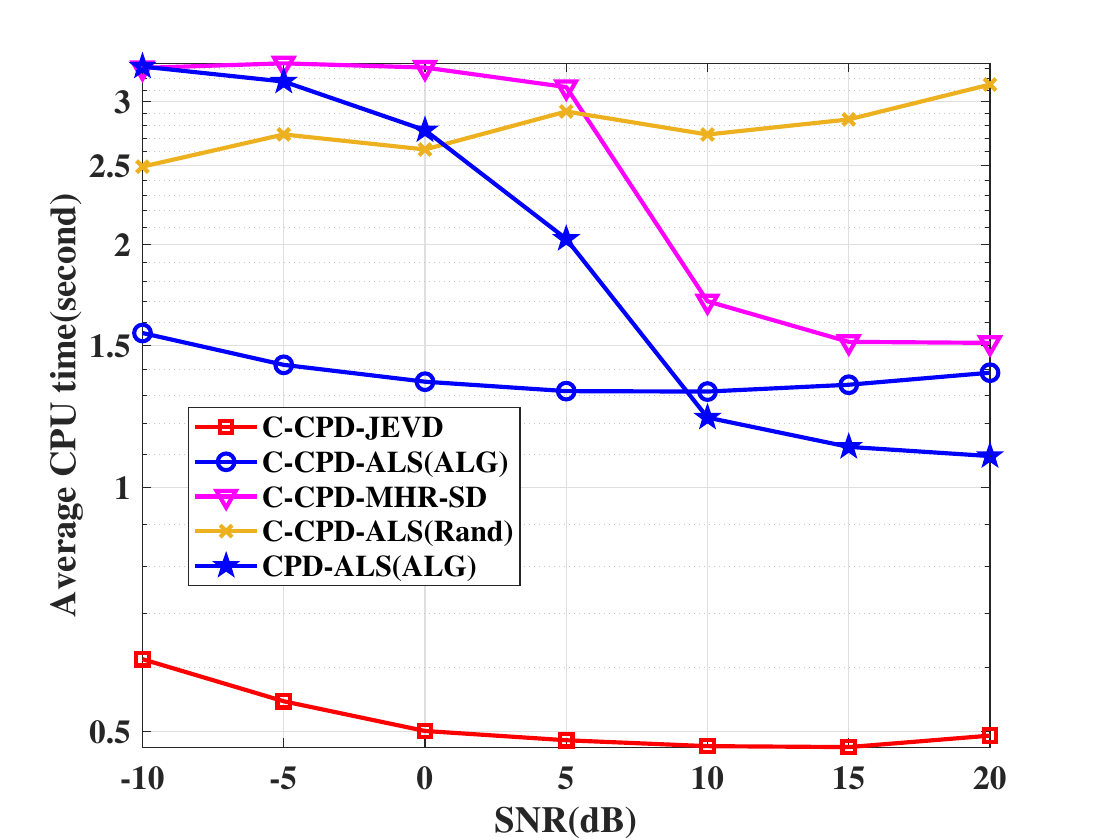}
		\caption{}
		\label{A_3_b}
	\end{subfigure}
	\caption{(a) Average MAE and (b) average CPU time vs. SNR in \textit{Experiment A-3}.}
	\label{A_3}
\end{figure}

Figs.~\ref{A_3_a} and~\ref{A_3_b} show the average MAE and CPU time versus SNR for the methods in \textit{Experiment A-3}. In this highly single underdetermined scenario, similar to Fig.~\ref{A_2}, the C-CPD-ALS(ALG) method shows the best performance in terms of MAE. \modified{The main difference between \textit{Experiment A-2} and \textit{Experiment A-3} is that as underdeterminedness increases, the performance gap between C-CPD-JEVD and C-CPD-MHR-SD, as shown in Fig.~\ref{A_3_a}, widens significantly compared to that in Fig.~\ref{A_2_a}, especially at low SNR values ranging from \(-10\) dB to \(10\) dB.} For optimization-based methods, C-CPD-ALS(ALG) leads, followed by C-CPD-ALS(Rand) and CPD-ALS(ALG). The CPD-ALS(ALG) method consistently produces suboptimal solutions across all SNR conditions in this scenario, while C-CPD-ALS(Rand) fails to converge within the maximum number of iterations due to random initialization. The CPU time results remain consistent with those observed in \textit{Experiments A-1} and \textit{A-2}, with C-CPD-JEVD again demonstrating the lowest computational cost.

Combining the results from \textit{Experiments A-1} to \textit{A-3}, the proposed methods consistently outperform the compared tensor-based algorithms in both MAE and CPU time. This is primarily due to efficient utilization of the Vandermonde structure in the C-CPD model \eqref{eq:Xm}, particularly the sparse ULA in the CPLsA. By transforming the C-CPD problem into a more efficient J-EVD problem, the C-CPD-JEVD method provides algebraic solutions that serve as low-cost initial values for the iterative SGSD algorithm, leading to noise-robust and accurate final estimates. Consequently, the proposed methods exhibit superior noise robustness and computational efficiency, making them highly effective for target localization in challenging scenarios.

\subsection{Experiments for MS CPPA}
 \label{sec:CaseB}
In this subsection, we evaluate the target localization performance of the compared algorithms in an MS CPPA MIMO radar system with the same array configuration. For the receive array, $M_x \!= \! M_y \!=\! 3, N_x \!=\! N_y \!=\! 5, I_{x,1}^{(m)} \!=\! I_{x,2}^{(m)} \!=\! I_{y,1}^{(m)} \!=\! I_{y,2}^{(m)} \!=\! 3$, and $I \!=\! 25$. For the transmit array, $M_x \!=\! M_y \!=\! 4, N_x \!=\! N_y \!=\! 7, J_{x,1} \!=\! J_{x,2} \!=\! J_{y,1} \!=\! J_{y,2} \!=\! 4$, and $J \!=\! 49$. Here, $J_{x,1}, J_{x,2}, J_{y,1}$, and $J_{y,2}$ are the number of sparse ULA elements along the \(x\)- and \(y\)-axes in the transmit array, corresponding to those in the \(m\)-th receive array.

We consider two scenarios: overdetermined and single underdetermined cases. In \textit{Experiment B-1}, \( I = 25 \), \( J = 49 \), \( K = 5 \), and \( R = 20 \), corresponding to an overdetermined case since \( \min(I, J) > R \). In \textit{Experiment B-2}, \( I = 25 \), \( J = 49 \), \( K = 15 \), and \( R = 30 \), representing a slightly single underdetermined case where \( I < R \). In \textit{Experiment B-3}, \( I = 25 \), \( J = 49 \), \( K = 45 \), and \( R = 45 \), defining a highly single underdetermined case, where \( J > R \) but \( R \approx 2I \). The SNR varies from -20 dB to 10 dB in \textit{Experiment B-1} and \textit{Experiment B-2}, and from -5 dB to 25 dB in \textit{Experiment B-3}. 

\begin{figure}[t]
 	\centering
 	\begin{subfigure}[]{0.45\textwidth}
 		\centering
 		\includegraphics[width=\textwidth]{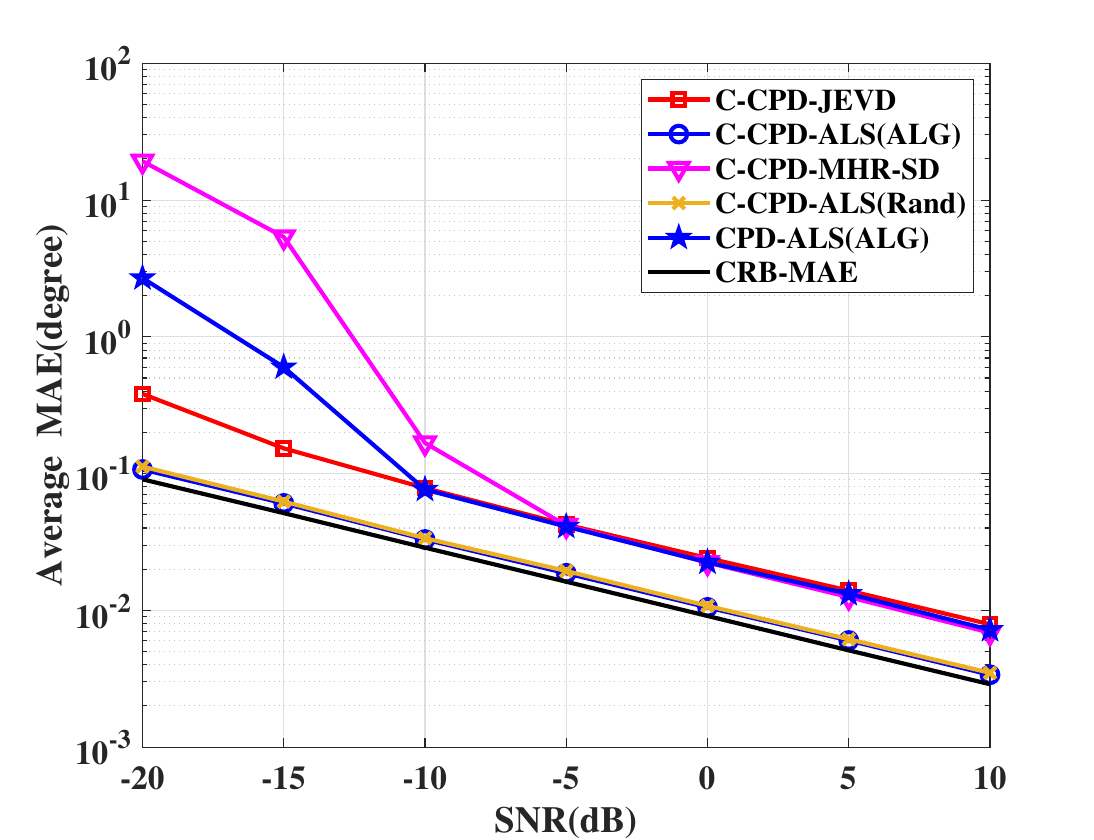}
 		\caption{}
 		\label{B_1_a}
 	\end{subfigure}
 	\begin{subfigure}[]{0.45\textwidth}
 		\centering
 		\includegraphics[width=\textwidth]{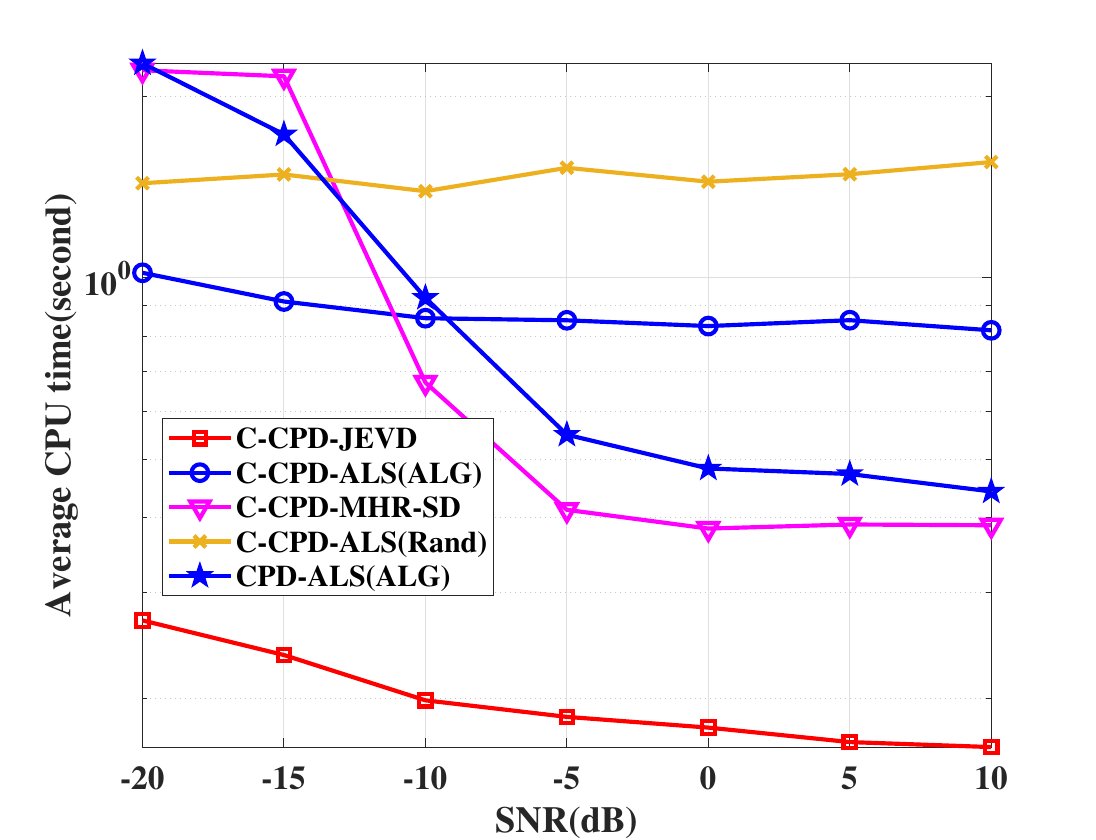}
 		\caption{}
 		\label{B_1_b}
 	\end{subfigure}
 	\caption{(a) Average MAE and (b) average CPU time vs. SNR in \textit{Experiment B-1}.}
 	\label{B_1}
 \end{figure}

Figs.~\ref{B_1_a} and~\ref{B_1_b} illustrate the average MAE and CPU time versus SNR for the methods compared in \textit{Experiment B-1}. In the overdetermined scenario (Fig.~\ref{B_1}), the proposed C-CPD-ALS(ALG) achieves the best MAE performance. Among semi-algebraic methods, C-CPD-JEVD outperforms C-CPD-MHR-SD, particularly at low SNRs (\(-20\) dB to \(-5\) dB). For optimization-based methods, C-CPD-ALS(ALG) leads, followed by C-CPD-ALS(Rand) and CPD-ALS(ALG), due to full exploitation of the Vandermonde and coupling structure within the MS CPLsA. Additionally, C-CPD-JEVD, C-CPD-MHR-SD, and CPD-ALS(ALG) exhibit similar MAE performance between 0 dB and 10 dB, reaching the expected performance limits of the C-CPD model \eqref{eq:Xm} in this well-conditioned scenario. Regarding CPU time, the proposed C-CPD-JEVD achieves the lowest computational cost by transforming the computationally and memory-intensive C-CPD problem into a more efficient J-EVD approach with reduced complexity and memory requirements. Furthermore, C-CPD-ALS(ALG), initialized with the solution from C-CPD-JEVD, converges faster than C-CPD-ALS(Rand), which uses random initialization.

\begin{figure}[t]
 	\centering
 	\begin{subfigure}[]{0.45\textwidth}
 		\centering
 		\includegraphics[width=\textwidth]{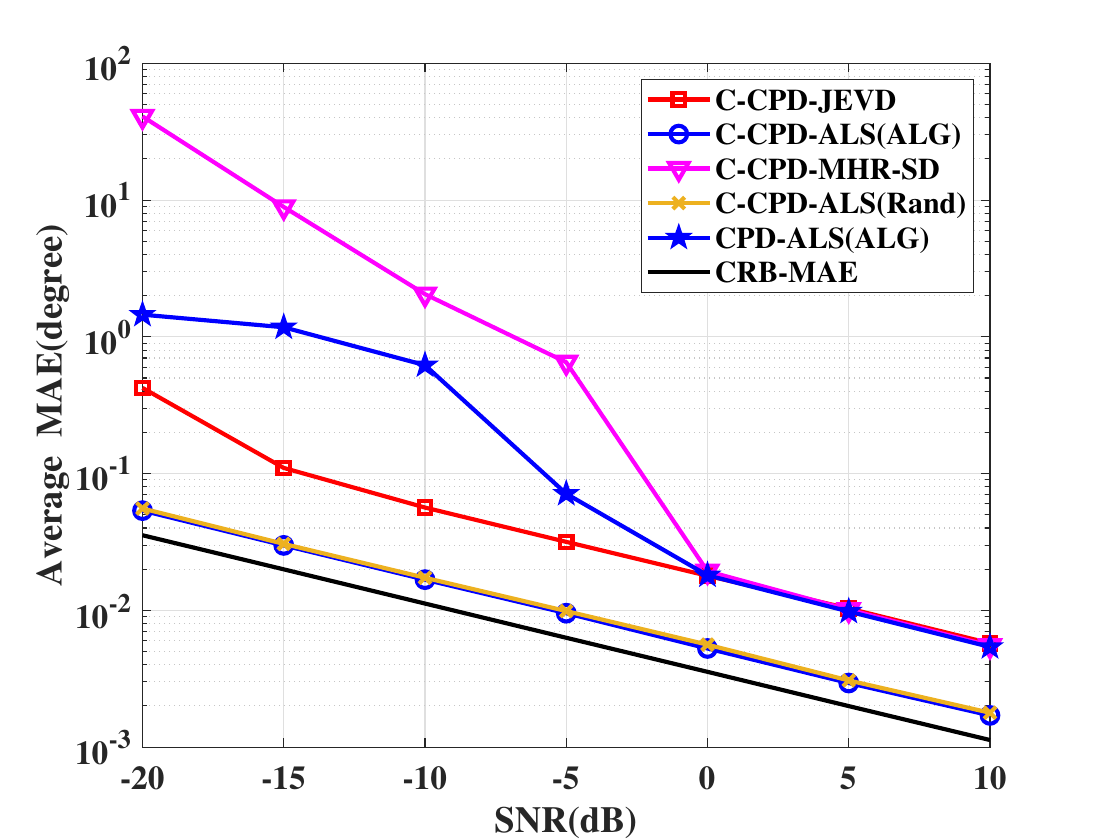}
 		\caption{}
 		\label{B_2_a}
 	\end{subfigure}
 	\begin{subfigure}[]{0.45\textwidth}
 		\centering
 		\includegraphics[width=\textwidth]{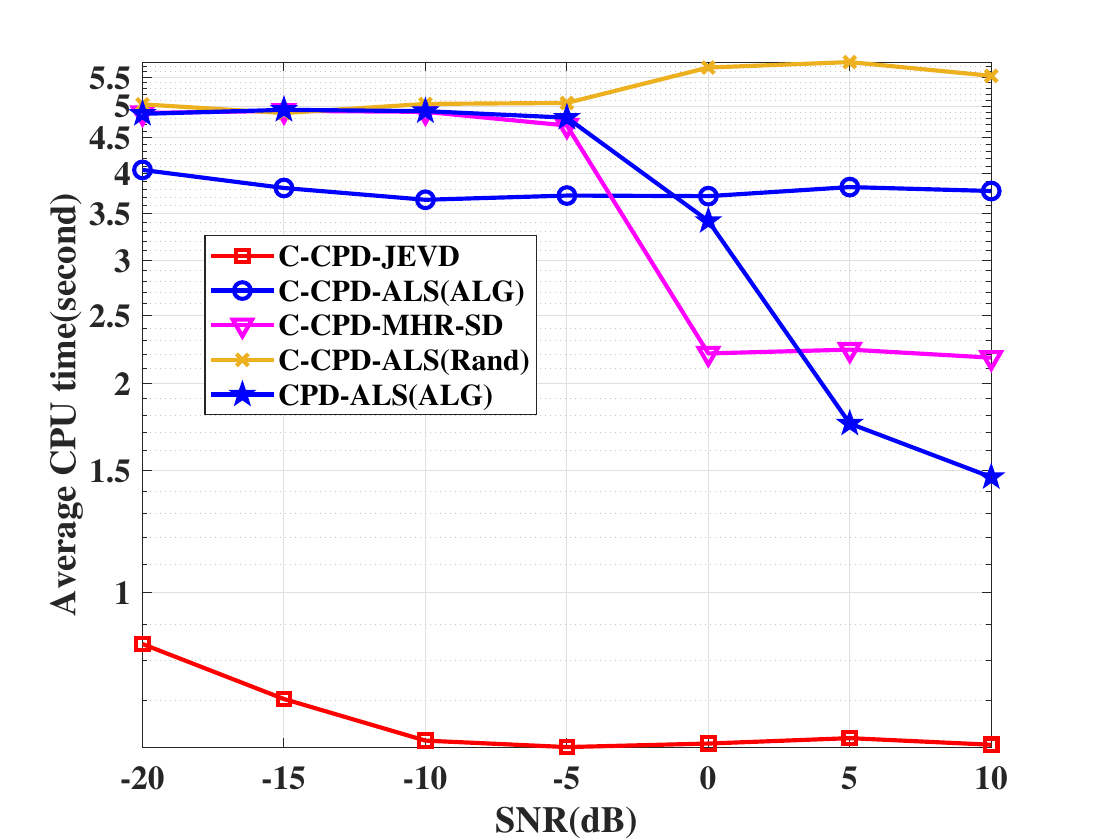}
 		\caption{}
 		\label{B_2_b}
 	\end{subfigure}
 	\caption{(a) Average MAE and (b) average CPU time vs. SNR in \textit{Experiment B-2}.}
 	\label{B_2}
 \end{figure}

Figs.~\ref{B_2_a} and~\ref{B_2_b} show the average MAE and CPU time versus SNR for the methods compared in \textit{Experiment B-2}. In the slightly single underdetermined case (Fig.~\ref{B_2}), similar to Fig.~\ref{B_1}, the proposed C-CPD-ALS(ALG) continues to demonstrate the best MAE performance. The key difference is that in Fig.~\ref{B_2_a}, the performance gap between the proposed C-CPD-JEVD and the SD-based C-CPD-MHR-SD widens significantly compared to Fig.~\ref{B_1_a}, particularly at low SNR values (\(-20\) dB to 0 dB) due to the superior noise robustness of the proposed methods compared to SD-based C-CPD-MHR-SD. In terms of CPU time, the results remain consistent with those of \textit{Experiment B-1}, with C-CPD-JEVD maintaining the lowest computational cost.
 
\begin{figure}[t]
	\centering
	\begin{subfigure}[]{0.45\textwidth}
		\centering
		\includegraphics[width=\textwidth]{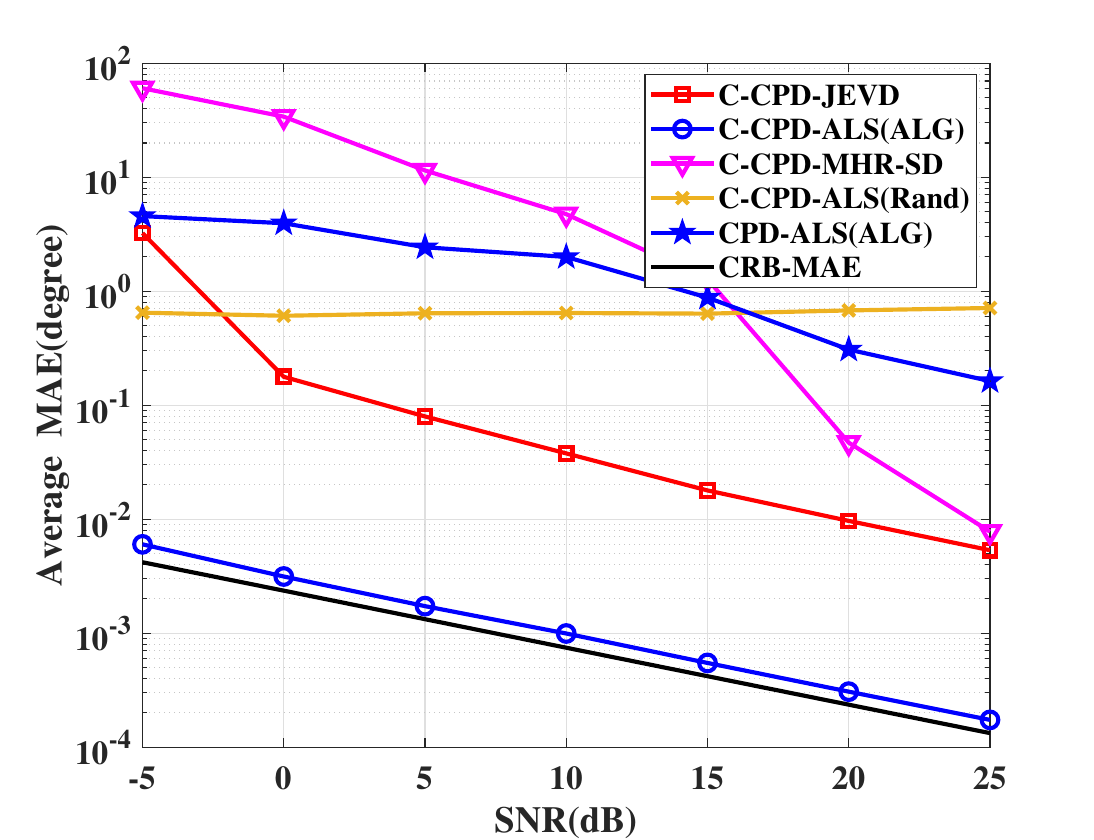}
		\caption{}
		\label{B_3_a}
	\end{subfigure}
	\begin{subfigure}[]{0.45\textwidth}
		\centering
		\includegraphics[width=\textwidth]{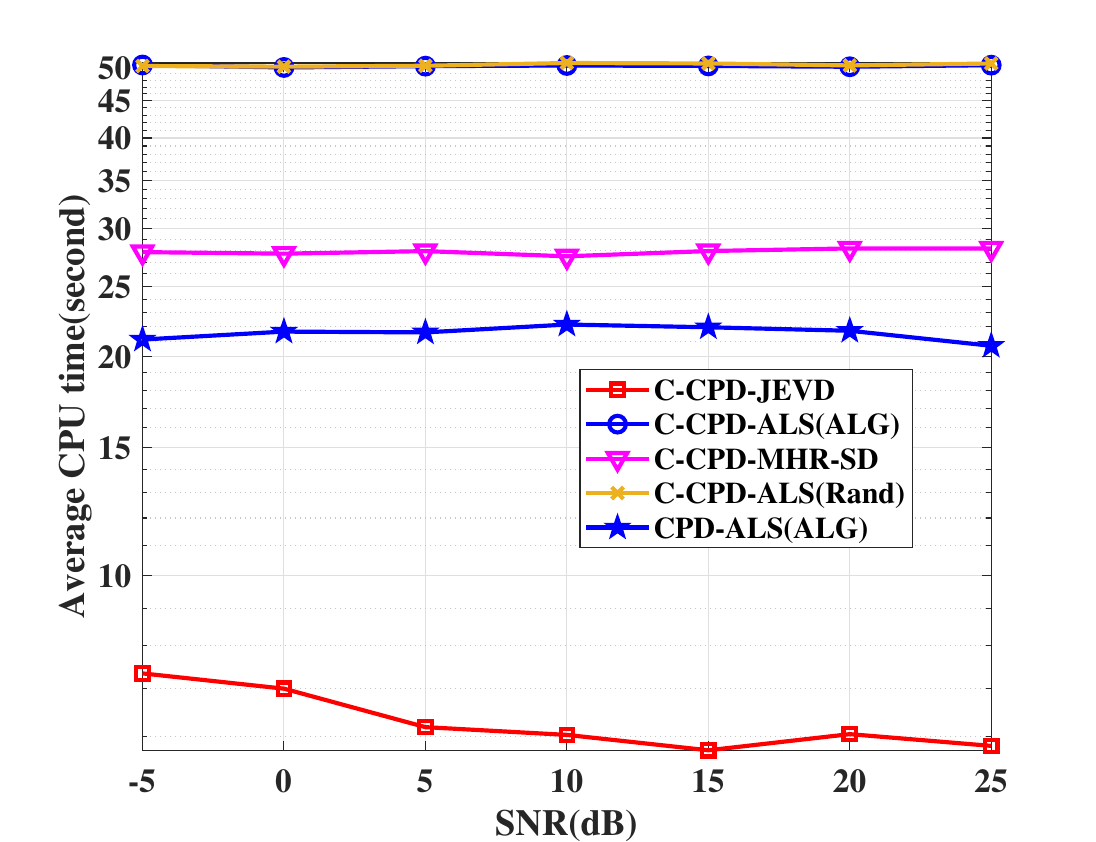}
		\caption{}
		\label{B_3_b}
	\end{subfigure}
	\caption{(a) Average MAE and (b) average CPU time vs. SNR in \textit{Experiment B-3}.}
	\label{B_3}
\end{figure}

 Figs.~\ref{B_3_a} and~\ref{B_3_b} show the MAE and CPU time versus SNR for the methods compared in \textit{Experiment B-3}. In this highly single underdetermined scenario, similar to Fig.~\ref{B_2}, the proposed C-CPD-ALS(ALG) method achieves the best MAE performance. The main difference is that as underdeterminedness increases, the performance gap between C-CPD-JEVD and C-CPD-MHR-SD widens significantly compared to Fig.~\ref{B_2_a}, especially at low SNR values ranging from \(-5\) dB to \(25\) dB. For optimization-based methods, only C-CPD-ALS(ALG) achieves ideal performance. The CPD-ALS(ALG) method consistently produces suboptimal solutions across all SNR conditions in this scenario, while C-CPD-ALS(Rand) fails to converge within the maximum number of iterations due to random initialization, resulting in poor performance. As with \textit{Experiment B-2}, the CPU time results remain the same as those in \textit{Experiment B-1}, confirming that C-CPD-JEVD still achieves the lowest computational cost.

In \textit{Experiments B-1} to \textit{B-3}, the proposed J-EVD-based semi-algebraic C-CPD framework consistently outperforms tensor-based methods in both MAE and computational cost, aligning with the results observed in \textit{Experiments A}. This superior performance is attributed to its full exploitation of the Vandermonde structure in the factor matrices of the C-CPD model \eqref{eq:Xm}, i.e., the sparse uniform array in the CPPA. By transforming the C-CPD problem into an efficient J-EVD problem, the C-CPD-JEVD method provides accurate initializations for the iterative SGSD algorithm, yielding robust and precise estimates. Consequently, the proposed approach achieves optimal performance in both overdetermined and single underdetermined scenarios.

\section{Conclusion}
\label{sec5}
A novel algorithmic framework has been proposed for target localization in MS MIMO radar systems, specifically designed for CPLsAs and CPPAs. For each array configuration, a semi-algebraic C-CPD algorithm based on J-EVD is developed, which reduces computational complexity by leveraging the rotational invariance and coupling structures inherent in MS coprime arrays. Additionally, initializing the iterative SGSD algorithm with GEVD enhances noise robustness. Working conditions of the J-EVD-based C-CPD algorithm are provided to ensure model identifiability, emphasizing that the conditions for the semi-algebraic algorithms associated with CPLsA and CPPA differ, thereby demonstrating the framework's applicability across various array geometries. The proposed method, which requires no prior knowledge of probing waveforms, achieves efficient target localization by fusing DOA information obtained from the J-EVD results. Experimental results show that the proposed method outperforms existing tensor-based approaches for both CPLsAs and CPPAs, yielding better MAE performance and lower computational cost in both overdetermined and underdetermined scenarios.
	
\section*{Acknowledgments}
		This work was supported in part by the National Natural Science Foundation of China under Grants 62471084 and 62071082, and in part by the China Postdoctoral Science Foundation under Grant 2020M680922.

\setlength{\bibsep}{0pt} 
\bibliographystyle{elsarticle-num}		
\bibliography{C_CPD_JEVD_bib_modified}

	\end{document}